\documentclass[preprint,showpacs,floatfix,12pt]{revtex4}

\usepackage{amsmath}
\usepackage{amsfonts}
\usepackage{amssymb}
\usepackage{bbm}
\usepackage{latexsym}
\usepackage[dvips]{graphicx}
\usepackage{color}

\begin{document}

\title{Equation of state for the Universe from similarity symmetries.}
\author{Marek Szyd{\l}owski, W{\l}odzimierz God{\l}owski,
Rados{\l}aw Wojtak}
\email{uoszydlo@cyf-kr.edu.pl}
\affiliation{Astronomical Observatory, Jagiellonian University,
Orla 171, 30-244 Krak{\'o}w, Poland}
\date{\today}
 
\begin{abstract}
In this paper we
proposed to use the group of analysis of symmetries of the dynamical system to
describe the evolution of the Universe. This methods is used in searching for
the unknown equation of state. It is shown that group of symmetries enforce
the form of the equation of state for noninteracting scaling multifluids.
We showed that symmetries give rise the equation of state in the form
$p=-\Lambda+w_{1}\rho(a)+w_{2}a^{\beta}+0$ and energy density
$\rho=\Lambda+\rho_{01}a^{-3(1+w)}+\rho_{02}a^{\beta}+\rho_{03}a^{-3}$,
which is commonly used in cosmology. The FRW model filled with
scaling fluid (called homological) is confronted with the observations of distant type
Ia supernovae. We found the class of model parameters admissible by the
statistical analysis of SNIa data. We showed that the model with scaling fluid
fits well to supernovae data. We found that $\Omega_{\text{m},0} \simeq 0.4$
and $n \simeq -1$ ($\beta = -3n$), which can correspond to (hyper) phantom
fluid, and to a high density universe. However if we assume  prior
that $\Omega_{\text{m},0}=0.3$ then the favoured model is close to concordance
$\Lambda$CDM model. Our results predict that in the considered model with
scaling fluids distant type Ia supernovae should be brighter than in
$\Lambda$CDM model, while intermediate distant SNIa should be fainter
than in $\Lambda$CDM model. We also investigate whether the model with
scaling fluid is actually preferred by data over $\Lambda$CDM model. As a
result we find from the Akaike model selection criterion prefers the model with
noninteracting scaling fluid.
\end{abstract}

\maketitle

\section{Introduction}
While the structure of space-time is governed by the field equations
the  physical properties of matter, introduced via energy-momentum tensor,
is take from other parts of physics.
 
Mathematically, one proceeds in an analogous way to that adopted by Newton
who defined a dual geometric-physical concept of mass-point. In General
Relativity one defines particles as curves in space-time and ascribes to
them physical quantities such as density and pressure. The $particle$ of the
rest-mass $m$ is defined as a future directed curve $\gamma\colon I\to M$ in a
space-time $M$, such that $g(\gamma_{\ast},\gamma_{\ast})=-m^{2}$. The tangent
vector $\gamma_{\ast}$ is the energy-momentum vector of the particle. Then,
a particle flow of rest-mass $m$ is defined to be a pair $(\vec{P},\eta)$
where $\eta$ is a function $\eta \colon M\to [0,\infty)$ called the world
density, and $\vec{P}$ is the energy-momentum vector field $\vec{P}\colon
M\to T(M)$ such that each integral curve of $\vec{P}$ is a particle of
rest-mass $m$. The energy-momentum tensor of particle flow $(\vec{P},\eta)$
is tensor $T=\eta\vec{P}\otimes \vec{P}$; and pressure $p$, understood as a
function of $M$, enters it through $\vec{P}$ (for detailed formalism see
\cite{Sachs77,Heller83}). Mc Crea \cite{McCrea51} noticed many years ago
that since matter is taken from outside of general relativity, there are no
a priori reasons why $p$ should be non--negative. The only way how $p$
should be interpreted is through effects which it produces in a model.
For example, $p<0$ may be interpreted as a quintessence matter
\cite{Peebles03} or dark energy due to the present Universe is accelerating
\cite{Perlmutter,Riess98}.
 
Lie groups of symmetry play an important role in searching for new
solutions but their original field of fruitful applications remained hidden
in the literature. This problem was discussed by Stephani \cite{Stephani89}
in his introduction to the basic monograph on the application of symmetry
groups to general relativity. But this method is still obscure because
many people who used them implicitly simply were not aware of their
existence \cite{Aug03a,Aug03b,Belinchon00,Belinchon02}.
 
The equation of state also plays an important role in general relativity.
As it is well known, Einstein's field equations, together with the Bianchi
identities, form an undetermined system of equations and one more equation,
equation of state must be added. Einstein's equations completed by the
equation of state satisfy certain symmetries which has structure of a group,
called a symmetry group (see \cite{Arnold80,Collins77a}). It is interesting
that this procedure can be reversed \cite{Collins77a}. Collins applied the
method of deriving form of the equation of state from symmetry to the case of
classical and relativistic stars to obtain the physically realistic equations
of state \cite{Collins77a,Collins77b}.
This illustrate--to use Collin's expressions--a subliminal role of
mathematics in physical considerations which consists in suggesting correct
physics on the ground of a logical beauty (such as, for example, symmetry
principles).
 
There are several interesting papers where the authors study symmetry
transformations under which the Einstein equation are invariant.
They relate symmetry and inflation \cite{Aug03a,Aug03b,Aug03c}
however the Lie group of symmetries is not used in analysis of the FRW equation.
The investigation of Lie symmetries of General Relativity and cosmology and
the origin of dynamical systems in the context of casual viscous fluid and
the cosmology with variable constants can also be found interesting.
For a cosmological models with bulk viscosity idea of renormalisation group
is also presented to study some scaling properties of the model
\cite{Belinchon00,Belinchon02}.
 
Because the Einstein equation constitute very complicated system of
nonlinear partial differential equation it is not easy to obtain an exact
solution. However, one can assume self-similarity symmetry and then
due to this simplified assumption Einstein field equation with spherical
symmetry can be reduced to the form ordinary differential equations i.e.
a dynamical system. Therefore the assumption of self-similarity symetry is
very useful in the problem of finding exact solutions in nonhomogeneous
case (see \cite{Carr99,Maeda04}). The self-similar solutions play especially
important roles in the cosmological applications or description of
gravitational collapse. It is the consequence of the fact that they
describe asymptotic behaviour of more general solutions. The idea that
spherical symmetric fluctuations might naturally evolve toward self-similar
form is known as as the similarity hypothesis  \cite{Carr93}.
 
A very interesting result was founded by Harada and Maeda \cite{Harada01}.
They proved numerically that self-similar solution acts as an attractor
in the phase space for spherically symmetric collapse of perfect fluid. 
In the context of gravitational collapse self-similar solution has been 
also studied from the points of view of critical  phenomena \cite{Choptuik93}.
 
The self-similar solutions also play an important role in the contemporary
context of dark energy. The scaling solutions in the context of "cosmic 
coincidence conundrum" were also considered in \cite{Sahni02}. It is 
demonstrated in many papers that there exists a stationary scaling solution 
in this problem which is representing a stable attractor at late times
\cite{Guo04,Maeda04,Tsujikawa04,Gumjudpai05,Amendola00}.
Among the class of solutions, the global atractor solutions are of the 
greatest interest from the physical point of view. They usually constitute 
scaling type of solutions which can explain why energy densities of dark 
energy and matter at present epoch are of the same order of magnitude 
\cite{Amendola99,Amendola04,Majerrotto04}.
 
Our main idea is to use the concept of self-similarity to the derivation of 
the form of equation of state. They reproduce themselves as the scale changes.
This methodology was already used by  Ellis and Buchert in context of problem 
of gravitational entropy \cite{Ellis05}.
 
To determine the structure and the dynamics of astrophysical systems as well
as the Universe, the equation  of state is usually necessary. Let us consider,
for example, the structure of a spherical neutron star. Then if the
pressure $p$ is known as a function of the density $\rho$, we can determine
the gravitational mass $M$ and the radius $R$ of the star as a function of
the central density $\rho_{c}$ by solving the Oppenheimer-Volkoff equation.
This means that we can determine in principle the mass-radius relation $M(R)$
theoretically. In similar way if the dependence between total pressure and
total energy density of the Universe is postulated then we can integrate the
Friedmann equation \cite{Harko04}. However, the equation of state relevant
to the neutron star or the Universe is not established yet, although it may
be determined in the near future. Let us note that for the Universe the form
of equation of state is not theoretically known at present but from
observations of SNIa data it may be reconstructed in the SNAP3.
\footnote{http://www-supernova.lbl.gov, http://snfactory.lbl.gov}.
 
In our paper the form of the equation of state is not postulated a priori but
it is derived from the existence of scaling solutions. The existence
of scaling solutions is equivalent the existence of a corresponding symmetry
group (similarity group) of the basic differential equations. It is just
discussed by us the type of symmetry of dynamical equations. Our philosophy is
to assume the existence of scaling solutions instead of postulated a priori
any prescribed form of the equation of state.
 
We use Lie symmetries to determine the dependence of energy density on the
scale factor. Of course if we assume the exact form of the equation of state
for example for phantoms then it is easy to obtain the exact form of
$\rho_{\text{eff}}(a)$ for the adiabatic condition.
 
In physical applications the important role is played by the scaling
solutions, they are also called self-similar, homological etc. The physical
properties of such solutions are important in many branches of physics
\cite{Falle91,Aug03c,Copeland98}. They can be found in the exact form or
can be analysed by using the dynamical system methods. Then invariants of
self-similarity groups are a good choice for a variable to parameterise
a system.
 
A natural question is whether the cosmological model filled by scaling fluid
is actually preferred by the data over concordance $\Lambda$CDM model. This
problem can be investigated by using the Akaike (AIC) and Bayesian (BIC)
informative criterion. We find that AIC favoured FRW model with scaling fluid
relative to $\Lambda$CDM model.
 
In the present paper we pursue Collins' way of thinking
\cite{Collins77a,Collins77b} applying it to the field of cosmology. In
section 2 we outlined the theory of symmetry groups of
differential equations, then we showed that cosmological equations, together
with a suitable equation of state admit a certain Lie group of symmetries or,
vice versa, the invariance of equations with respect to a given symmetry
group singles out corresponding equation of state. In sections 3-4 we showed
this for Friedmann models with matter in the form of perfect fluid and
effective cosmological constant, respectively. In section 5 and 6 we
confronted data obtained theoretically from symmetry equation of state with
recently available SNIa data. In section 7 we briefly commented on the
obtained results.

\section{Symmetry group of a system of differential equations.}
 
Let us consider, in a Euclidean space $E^{N}(x,u)$,
$x=(x^{1},\dots,x^{n}), u=(u^{1},\dots,u^{m})$,
$n+m=N$, the system of differential equations
\begin{equation}
p^{\alpha}=f^{\alpha}(x,u),\qquad \alpha=(1,\dots,m),\qquad
p^{\alpha}=\frac{\partial u^{\alpha}}{\partial x}
\label{eq:1}
\end{equation}
and point-point transformations
\begin{equation}
T: \left\{ \begin{array}{ll}
   \bar{x}=\bar{x}(x,u),\\
   \bar{u}^{\alpha}=\bar{u}^{\alpha}(x,u),
\end{array} \right.
\label{eq:2}
\end{equation}
which map each solution of system (\ref{eq:1})
into a solution of the same system. $T$ is a Lie group. (All these
considerations may be easily generalised to the
case when differential equations are defined on a $N$-dimensional
differential manifold; see \cite{Stephani89,Hydon99}).
 
Let $M\colon u=u^{\alpha}(x)$ be a solution of (\ref{eq:1}).
$M$ defines a submanifold in $E^{N}$; if $\bar{M}\equiv T(M)$ we have
\begin{equation}
\overline{M}: \left\{ \begin{array}{ll}
       \bar{x}=\bar{x}(x,u(x)),\\
       \bar{u}^{\alpha}=\bar{u}^{\alpha}(x,u(x)),
\end{array} \right.
\label{eq:3}
\end{equation}
The derivatives $p^{\alpha}=\frac{\partial u^{\alpha}}{\partial x}$ and
$\bar{p}^{\alpha}=\frac{\partial \bar{u}^{\alpha}}{\partial \bar{x}}$
satisfy the condition
\begin{equation}
\bar{p}^{\alpha}D\bar{x}=D\bar{u}^{\alpha},
\label{eq:4}
\end{equation}
where
$D=\frac{\partial}{\partial x}+p^{\alpha}\frac{\partial}{\partial u^{\alpha}}$.
By solving (\ref{eq:4}) we obtain $\bar{p}^{\alpha}=\bar{p}^{\alpha}(x,u,p)$.
By joining these solutions to transformations $T$, one gets a new set of
transformations $\tilde{T}$ which is called ``extension of $T$ to the first
derivatives''.
 
If the infinitesimal operator of the group $T$ has the form
\begin{equation}
X=\xi(x,u)\frac{\partial}{\partial x}
+\eta^{\alpha}(x,u)\frac{\partial}{\partial u^{\alpha}}
\end{equation}
then that the group $\tilde{T}$ is
\begin{equation}
\tilde{X}=X+(D\eta^{\beta}-p^{\beta}D\xi)\frac{\partial}{\partial p^{\beta}}
\label{eq:5a}
\end{equation}
Function $\mathcal{F}$ defined on $\tilde{E}^{N}(x,u,p)$, which are preserved
under transformations $\tilde{T}$, are
called invariants of the group $T$; they satisfy differential equations
\begin{equation}
\tilde{X}(\mathcal{F})=0
\label{eq:5}
\end{equation}
and, vice versa, functions satisfying (\ref{eq:5}) are invariants of
the group generated by $X$.
 
Now, we may apply equations (\ref{eq:5}) to our original equations (\ref{eq:1})
to obtain
\begin{equation}
\tilde{X}(p^{\alpha}-f^{\alpha}(x,u))=0
\label{eq:6}
\end{equation}
Putting (\ref{eq:5a}) into (\ref{eq:6}) and taking into account $p^{\alpha}=f^{\alpha}(x,u)$, we get
\begin{equation}
\frac{\partial \eta^{\alpha}}{\partial x}+\frac{\partial \eta^{\alpha}}{\partial u^{\beta}}f^{\beta}-
\frac{\partial \xi}{\partial x}f^{\alpha}-\frac{\partial \xi}{\partial u^{\beta}}f^{\beta}f^{\alpha}
=\xi\frac{\partial f^{\alpha}}{\partial x}+\eta^{\beta}\frac{\partial f^{\alpha}}{\partial u^{\beta}},\qquad
\alpha,\beta=(1,\dots,m).
\label{eq:6a}
\end{equation}

\section{The Group  Symmetry of Einstein equations with the perfect fluid energy-momentum tensor.}
 
The problem of group properties of vacuum Einstein equations has been
investigated by Ibragimov (1983) \cite{Ibragimov83}. In this case we obtain
one parameter group of rescalings of the metric $g_{ij}\to \alpha g_{ij}$
together with infinite group of coordinate transformations from the maximal
group of point transformations admitted by  vacuum Einstein equations
$R_{ik}=0; i,k=0,1,2,3$. We shall investigate symmetries of Einstein equations:
\begin{equation}
R_{ik}=T_{ik}-\frac{1}{2}Tg_{ik} \qquad i,k=0,1,2,3
\label{eq:7}
\end{equation}
We assume that $T_{ik}$ is the perfect fluid energy-momentum tensor.
We choose the comoving coordinate system i.e.
$u_{i}=\delta_{0i}$.\\
Then we have:
\begin{eqnarray}
T_{ik}=(\rho+p)\delta_{0i}\delta_{0k}-pg_{ik} \nonumber\\
T=\rho-3p,
\label{eq:8}
\end{eqnarray}
where $p=p(\rho)$.\\
We assume that Einstein equations (\ref{eq:7}) with energy-momentum tensor
(\ref{eq:8}) possesses the same symmetries as corresponding  vacuum
equations. We shall look the form of equation of state enforced by symmetries
of the vacuum Einstein's equations. Infinitesimal operator of the Lie group
 symmetry acting in the set of Einstein equations can be written in the form:
\begin{equation}
\tilde{X}=\xi^{i}(x)\frac{\partial}{\partial x^{i}}+\eta_{ij}(x,\rho)\frac{\partial}{\partial g_{ij}}+\theta(x,g,\rho)
\frac{\partial}{\partial \rho}.
\label{eq:9}
\end{equation}
where $\theta$ can be derived from the condition for operator
(\ref{eq:9}) to be admissible operator of equation (\ref{eq:6a}).
This condition is the following:
\begin{equation}
\tilde{X}_{2}[R_{ik}-T_{ik}+\frac{1}{2}Tg_{ik}]=0.
\end{equation}
The operator $\tilde{X_{2}}$ is the extension of $\tilde{X}$ to the second derivatives:
\begin{align}
\tilde{X}_{2} &= X_{2}+\theta\frac{\partial}{\partial \rho} \nonumber\\
\tilde{X}_{2} &= \tilde{X}+\xi_{ijk}\frac{\partial}{\partial g_{ij,k}}
+\xi_{ijkl}\frac{\partial}{\partial g_{ij,kl}}\nonumber\\
&\tilde{X}_{2}[R_{ik}]=0
\end{align}
It was demonstrated that symmetries of the vacuum Einstein equations enforce
the equation of state $p=\text{const}+w\rho$ ($w=\text{const}$) if we
postulate that the source of gravity is given in the form of perfect fluid
\cite{Biesiada89}.

\section{Symmetry group of Friedmann equations}
 
Equation~(\ref{eq:6a}) gives us the conditions of the existence of the operator
symmetry $X$ for the system (\ref{eq:1}). Algebra of these operators
characterises the symmetry of the group $T$ symmetry of system~(\ref{eq:1}).
 
In the following part  we shall consider symmetry of autonomous dynamical systems for
which $\dot{x}^{\alpha}=f^{\alpha}(x)$ (a dot denotes the differentiation with
respect to the parameter $u$), $\frac{\partial f^{\alpha}}{\partial x}=0$ and
infinitesimal transformations are generated by
$X=\xi(x)\frac{\partial}{\partial x}+\eta^{\alpha}(u)\frac{\partial}{\partial
u^{\alpha}}$. In this case (\ref{eq:6a}) simplify to the form
\begin{equation}
\frac{\partial \eta^{\alpha}}{\partial u^{\beta}}f^{\beta}-\frac{\partial \xi}{\partial x}f^{\alpha}=\eta^{\beta}
\frac{\partial f^{\alpha}}{\partial u^{\beta}}
\label{eq:10a}
\end{equation}
The Friedmann equations may be written in the form of
autonomous dynamical system on the plane $(a,\rho)$
\begin{eqnarray}
\dot{a}\equiv \frac{da}{dt}=\sqrt{\frac{\rho a^{2}}{3}-k+\frac{\Lambda a^{2}}{3}}=f^{1}(a,\rho) \nonumber\\
\dot{\rho}\equiv \frac{d\rho}{dt}=-\frac{3}{a}\sqrt{\frac{\rho a^{2}}{3}-k+\frac{\Lambda a^{2}}{3}}(\rho+p(\rho,a))=f^{2}(a,\rho)
\label{eq:10}
\end{eqnarray}
where $a$-- is a scale factor, $\rho$-- energy density, $k$-- curvature index,
$\Lambda$-- cosmological constant.
 
The set of basic dynamical equations is called the Friedmann dynamical system.
If we assume that the operator of infinitesimal transformation of symmetry
for system~(\ref{eq:10}) is of the form
\begin{equation}
X=\xi(t)\frac{\partial}{\partial t}+\eta^{1}(a)\frac{\partial}{\partial a}+\eta^{2}(\rho)\frac{\partial}{\partial \rho}
\end{equation}
then equations (\ref{eq:10a}) can be reduced to
\begin{eqnarray}
\frac{d\eta^{1}}{da}(a)-\frac{d\xi}{dt}=X(\ln f^{1})=\frac{\eta^{1}}{f^{1}}\frac{\partial f^{1}}{\partial a}+
\frac{\eta^{2}}{f^{1}}\frac{\partial f^{1}}{\partial \rho} \nonumber\\
\frac{d\eta^{2}}{d\rho}(\rho)-\frac{d\xi}{dt}=X(\ln f^{2})=\frac{\eta^{1}}{f^{2}}\frac{\partial f^{2}}{\partial a}+
\frac{\eta^{2}}{f^{2}}\frac{\partial f^{2}}{\partial \rho}
\label{eq:11a}
\end{eqnarray}
Let us concentrate now (without lose of degree of generalisation) on the
special case of the FRW, namely the flat model without cosmological constant.
Then equations~(\ref{eq:10}) reduces to
\begin{eqnarray}
\frac{da}{dt}=aE \nonumber\\
\frac{dE}{dt}=-\frac{1}{2}(3E^{2}+p(a,E))
\label{eq:11}
\end{eqnarray}
where $E=\sqrt{\rho /3}$ is redefined energy density, $p$ is a pressure as an
unknown function of scale factor. Let us assume that equations (\ref{eq:11})
admit the special symmetry of the operator for quasi-homology,i.e.
\begin{equation}
X=\xi(t)\frac{\partial}{\partial t}+\eta^{1}(a)\frac{\partial}{\partial a}+\eta^{2}(E)\frac{\partial}{\partial E}
\end{equation}
Then from equations (\ref{eq:11a}) we obtain
\begin{equation}
\frac{d \eta^{1}(a)}{da}-\frac{d \xi(t)}{dt}=\frac{\eta^{1}(a)}{a}+\frac{\eta^{2}(E)}{E}
\label{eq:12a}
\end{equation}
\begin{equation}
\frac{d \eta^{2}(E)}{dE}-\frac{d \xi}{dt}=\frac{1}{3E^{2}+p(a,E)}\Big(\eta^{1}(a)\frac{\partial}{\partial a}+
(6E+\frac{\partial p}{\partial E})\eta^{2}(E)\Big)
\label{eq:12b}
\end{equation}
>From (\ref{eq:12a}) we have that
\begin{equation}
\eta^{1}=Aa,\qquad \eta^{2}=DE,\qquad \xi(t)=Bt
\label{eq:13}
\end{equation}
where $A,D,B$ are constants which satisfying the constraint condition:
\begin{equation}
D=-B
\label{eq:14}
\end{equation}
After substitution (\ref{eq:13}) and (\ref{eq:14}) into the equation
(\ref{eq:12b}) we obtain  partial differential equation determining
$p=p(a,E)$
\begin{equation}
2Dp=Aa\frac{\partial p}{\partial a}+DE\frac{\partial p}{\partial E}
\label{eq:15}
\end{equation}
The solutions of equation (\ref{eq:15}) determine the form of the equation of
state required by the condition of quasi-similarity (quasi-homology).
The condition for the existence of quasi-homological type of symmetry reduces
to corresponding condition for existence of homological symmetry and then $p$
must satisfy (\ref{eq:15}).
 
Let us consider now some special solutions of (\ref{eq:15}).
If $p=p(E)$ only then $p\propto E^{2}=\rho$, while if $p=p(a)$ only then
$p(a) \propto a^{\frac{2D}{A}}$. It is easy to find the general solution of
(\ref{eq:15}) by using the standard characteristic methods. Then we obtain
\begin{equation}
\phi=\phi(J_{1},J_{2})
=\phi \left(\frac{E^{2}}{p},\frac{a^{\frac{2D}{A}}}{p}\right)=0
\label{eq:16}
\end{equation}
is the solution of eq.~(\ref{eq:15}), where $\phi \in \mathbb{C}^{1}$, and
$J_{1},J_{2}$ are its invariants.
 
There is equivalent form of the solution (\ref{eq:16}) in which the pressure
can be given in the exact form, namely
\begin{equation}
p=E^{2}g \left( \frac{a^{\frac{2D}{A}}}{E^{2}} \right)
\label{eq:17}
\end{equation}
or $g(x)\propto x$ where $g$ is any function of class $\mathbb{C}^{1}$.
If we substitute $g(x)=\text{const}$ in eq.~(\ref{eq:17}), then mentioned
before special cases of $p$ can be simply recovered. Therefore eq.~(\ref{eq:17})
gives the most general form of the equation of state for the Universe to be
self invariant or homological. In this case we can find a strict analogy to
stars \cite{Collins77a,Collins77b}. One should notice that the formulae
(\ref{eq:17}) contains for example the following form of the equation of state
\begin{equation}
p=-\Lambda+C_{1}\rho(a)+C_{2}a^{\frac{2D}{A}}
\label{eq:18}
\end{equation}
which is crucial for our further investigations.
 
We can find that eq.~(\ref{eq:18}) describes the pressure of noninteracting
multifluids components. Then from the conservation condition postulating for
each component we obtain
$\rho=\Lambda+\rho_{01}a^{-3(1+C_{1})}+\rho_{03}a^{-3}+\rho_{02}a^{\frac{2D}{A}}$,
where $\Lambda$, $\rho_{01}$, $\rho_{02}$, $\rho_{03}$ are constants.
Note that in the special case of $\frac{2D}{A}=-3$ equation (\ref{eq:18})
and conservation condition give rise to new type of contribution
$-(3\rho_{02} \ln a)a^{-3}$ instead $\rho_{02}a^{\frac{2D}{A}}$.
It corresponds to MOND phase squeezing in FRW scenario \cite{starkman}.
To obtain the form of equation of state (\ref{eq:18}) it is sufficiently to
choose $g(x)=-3+C_{1}+C_{2}x$. Then from equations~(\ref{eq:13}) the finite
transformations can be simply obtain for this case, namely
\begin{align}
a &\to \bar{a}=ae^{A\tau} \nonumber\\
E &\to \bar{E}=Ee^{D\tau} \nonumber\\
t &\to \bar{t}=te^{-D\tau},\qquad \tau \in \mathbb{R}^{1}
\end{align}
One can check that the invariant $a^{-\frac{2D}{A}}p$ is really preserved
under the action of the homological transformation, i.e.
\begin{equation}
a^{\frac{2D}{A}}p(a,E)=\overline{a}^{\frac{2D}{A}}p(\overline{a},\overline{E})
\end{equation}
The equation of state given by~(\ref{eq:17}) assumes  very general form.
Their existence
can be closely associated with the real equation of state for quintessence
matter. Let us note that equation~(\ref{eq:18}) contains not only standard term
of the equation of state $p \propto \rho$, but also the cosmological constant
with a constant part and a part depending on the scale factor $a$.
In the special case if $D=-A$ we recover the form of the effective cosmological
constant distinguished by a maximum of the amplitude for the tunnelling Gamov
process \cite{Jafarizadek99}. In the general case in the parameterisation of
the cosmological term it is usually assumed the power law form of $\Lambda(a)$
dependence which is also suggested from symmetry consideration
\cite{Silveira97}.
 
Note that homological transformations are strictly related to dimensional
analysis. The existence of such kind symmetry means that equation can be
simply expressed in the dimensionless form.  Therefore for each component
of the fluid one can define dimensionless density parameters which contain
basic Friedmann equation.

\section{Equation of state parameter from distant supernovae observations.}

In this section we confront FRW cosmological model filled with homological
fluid (which equation of state was previously suggested from symmetry arguments)
with observations of distant supernovae type Ia. These interpreted observations
in the framework FRW model indicate that our universe is presently accelerating
due to unknown form of matter called dark energy with negative pressure
\cite{Alam03,Gorini04}.There are different candidates for dark energy which was tested
from SNIa observations
\cite{Godlowski03,Godlowski04,Godlowski04a,Dabrowski04,Godlowski04b,Biesiada05}
 
The most popular candidate is cosmological
constant which can be treated as a perfect fluid for which $p=-\rho$ and
$\rho=\Lambda$. Although cosmological constant explains observations of SNIa
we don't understand why the observed value of cosmological constant
($\frac{\Lambda}{8\pi G}\sim 10^{-47}GeV^{4}$) is so small in comparison
with its natural theoretical expectations from quantum field
theory $\sim 10^{71}GeV^{4}$. According to this theory,the energy--momentum
tensor of the vacuum is nonvanishing and $<T_{\mu \nu}>=<\rho>g_{\mu \nu}$.
Therefore the observed cosmological constant term is
$\Lambda=\lambda+8\pi G<\rho>$ where $\lambda$ is the ``bare'' cosmological
constant. Our naive intuitions is that
$<\rho>\approx M_{\text{Pl}}^{4}$ while upper bound on the present value of
$\Lambda$ (referred as $\Lambda_{0}$) can be given in term of Hubble's
constant $H_{0}$
\begin{center}
$|\Lambda_{0}|/8\pi G \leq 10^{-29} g/cm^{3} \simeq 10^{-47} GeV^{4}$
\end{center}
The idea that cosmological constant is the sum of two terms $\lambda$ and
$8\pi G\rho_{vac}$ is proposed as a method of explaining the observed small
value of $\Lambda$ which might had been large in the early universe. The
variable $\Lambda$ models (or a decaying vacuum energy density) are described
in terms of two fluids mixture just like it was predicted from similarity
analysis. In the majority of the paper's calculations $\rho_{V}$ depends
only on the cosmological time through the scale factor $\rho_{V}\sim a^{-2}$.
The expression $\rho(a)$ is obtained by using dimensional arguments made in
the spirit of quantum gravity \cite{Silveira97} . We assume that we have
a decaying vacuum medium parameterised by scale factor in the power of
law plus a noninteracting cosmological constant term $\Lambda$. Additionally,
we usually  have baryonic matter satisfying the equation of state $p=0$.
Hence the FRW equation in general case can be rearranged to the form giving
the Hubble function $H(z)= \dot{a}/a$
 
\begin{equation}
\frac{H^2}{H_0^2}=\Omega_{\mathrm{m},0}\left(\frac{a}{a_0}\right)^{-3}
+\Omega_{\mathrm{C},0}\left(\frac{a}{a_0}\right)^{-3n}
+\Omega_{k,0}\left(\frac{a}{a_0}\right)^{-2}
+\Omega_{w,0}\left(\frac{a}{a_0}\right)^{-3 \left(1+w \right)}
+\Omega_{\Lambda,0}
\label{eq:101}
\end{equation}
 
For $a=a_0$ (the present value of scale factor) we obtain the following
constraint
\begin{equation}
\Omega_{\mathrm{m},0} + \Omega_{\mathrm{C},0} + \Omega_{k,0}
+\Omega_{w,0}+\Omega_{\Lambda,0}=1.
\label{eq:102}
\end{equation}
 
Because of the constraint $H^2 \ge 0$ the space admissible for the motion
should be restricted to the region at which $\rho_{\text{eff}}(a) \ge 0$.
If some negative component of effective energy density appears in
$\rho_{\text{eff}}(a)$ then it can never dominate at the early universe
and at the late times. If negative energy density scaling like $a^{-m}$
appears in such a way that
$3H^2=\Omega_{m,0}a^{-m}-\Omega_{n,0}a^{-n} + \Omega_{\Lambda,0}$ (where
$n>m$) then we have the case of bouncing cosmological models
\cite{Molina99,Tippett04,Szydlowski05}. In this model the relation
$\rho_{\text{eff}}(a) \ge 0$ is of course satisfied. In general the violation
of the strong energy condition (SEC) is a necessary (but not sufficient)
condition for the bounce to appear. The bouncing models can be characterised
by the minimal condition under which the present Universe arises from a bounce
with a previous collapse phase (note the reflectional symmetry of the basic
equation $H \to -H$) \cite{Molina99,Tippett04}. In order if we have such
a situation that $n<m$, then negative term $-\Omega_{n,0}a^{-n}$ cannot
dominate late evolution and therefore the scale factor should be bounded
and an inverted bounce appears.
 
If the energy density is very large then quantum gravity corrections are
important at both the big-bang and big-rip singularities. The account of
quantum effects leads to avoid not only the initial singularity
\cite{Bojowald01,Bojowald02} but also escape from the future singularity
\cite{Nojiri04a,Nojiri04b,Nojiri04c}.
 
The idea of bounce in FRW cosmologies appeared in Tolman's monograph devoted
to cosmology \cite{Tolman:1934}. This idea was strictly connected with
oscillating models \cite{Robertson:1933,Einstein:1931,Tolman:1931}.
At present oscillating models play an important role in the brane cosmology
\cite{Steinhardt:2002kw,Shtanov:2002ek}. The FRW universe undergoing a bounce
instead of the big-bang is also an appealing idea in the context of quantum
cosmology \cite{Coule:2004qf}. The attractiveness of bouncing models comes
from the fact that they have no horizon problem and they explain quantum
origin of structures in the Universe
\cite{PintoNeto:2004wf,Barrow:2004ad,Salim:2005}.
Molina-Paris and Visser and later Tippett \cite{Molina99,Tippett04}
characterised the bouncing models by the minimal condition under which the
present universe arises from a bounce from the previous collapse phase
(the Tolman wormhole is different name for denoting such a type of evolution).
The violation of strong energy condition (SEC) is in general a necessary (but
not sufficient) condition for bounce to appear. For closed models it is
sufficient condition and none of other energy condition need to be violated
(like null energy condition (NEC): $\rho+p \ge 0$, week energy condition
(WEC): $\rho \ge 0$ and $\rho + p \ge 0$, dominant energy condition (DEC):
$\rho \ge 0$ and $\rho \pm p \ge 0$ energy conditions can be satisfied).
 
We can find necessary and sufficient conditions for an evolutional path with
a bounce by analysing dynamics on the phase plane ($a, \dot{a}$), where $a$ is
the scale factor and dot denotes differentiation with respect to cosmological
time. We understand the bounce as in \cite{Molina99,Tippett04},
namely there must be some moment say $t=t_{\text{bounce}}$ in evolution of
the universe at which the size of the universe has a minimum,
$\dot{a}_{\text{bounce}}=0$ and $\ddot{a} \ge 0$. This weak inequality
$\ddot{a}\ge 0$ is enough for giving domains in the phase space occupied by
trajectories with the bounce.
 
We understand the inverted bounce as the situation in which in some moment
$t=t_{\text{bounce}}$ in evolution of the universe the size of the universe
has a maximum, $\dot{a}_{\text{bounce}}=0$ and $\ddot{a}_{\text{bounce}} \le 0$.
 
For our aim we considered some version above model in which for simplicity
we put $\Omega_{w,0}$. If we rewrite this equation in $z$ variable we obtain
for the flat model that
\begin{equation}
H(z)=H_{0}\sqrt{\Omega_{m,0}(1+z)^{3}+\Omega_{\lambda,0}
+(1-\Omega_{m,0}-\Omega_{\lambda,0})(1+z)^{\gamma}},
\label{eq:103}
\end{equation}
where $\gamma=3n$.
Our further task is be to confront the ``homological gas'' or variable
$\Lambda$ fluid with SNIa data and for this purpose we calculate the
luminosity distance in a standard way
\begin{equation}
d_{L}(z)=(1+z)\int_{0}^{z} \frac{d\bar{z}}{H(\bar{z})}
\label{eq:104}
\end{equation}
Further in this section we will use the flat FRW model since the evidence for
this case is very strong in the light of WMAP data \cite{Bennett03}. Therefore
while talking about model testing we actually mean the estimation of both
$\Omega_{\lambda,0}$ and $\gamma$ parameters for the best fitted flat model
with homological gas. Specifically we have tested this cosmology
with prior assumption $\Omega_{m,0}=0.3$ and $\Omega_{m,0}=0.05$ as well as
a model without prior assumption on $\Omega _{m,0}$.
 
To proceeded with fitting SNIa data we need magnitude-redshift relation
\begin{equation}
m(z,\mathcal{M},\Omega_{m,0};\Omega_{\lambda,0},\gamma)=\mathcal{M}+5\log_{10}D_{L}(z,\Omega_{m,0};\Omega_{\lambda,0},\gamma)
\label{eq:105}
\end{equation}
where:
\begin{equation}
D_{L}(z,\Omega_{m,0};\Omega_{\lambda,0},\gamma)=H_{0}d_{L}(z,H_{0},\Omega_{m,0};\Omega_{\lambda,0},\gamma)
\label{eq:105a}
\end{equation}
is the luminosity distance with $H_{0}$ factored out, so that marginalisation
over the intercept
\begin{equation}
\mathcal{M}=M-5\log _{10}H_{0}+25
\label{eq:106}
\end{equation}
leads actually to joint marginalisation $H_{0}$ and $M$ ($M$ being the absolute
magnitude of SNIa).

Then we can obtain the best fit model minimising the function $\chi^{2}$
\begin{equation}
\chi^{2}=\sum_{i}\frac{(m_{i}^{theor}-m_{i}^{obs})^{2}}{\sigma_{i}^{2}}
\label{eq:107}
\end{equation}
where their sum is over the SNIa sample and $\sigma_{i}$ denote the (full)
statistical error of magnitude determination. This is illustrated by figures
of residuals (with respect to Einstein--de Sitter model).
One of the advantages of residual
plots is that the intercept of the $m$--$z$ curve gets cancelled. The
assumption that the intercept is the same for different cosmological models
is legitimate since $\mathcal{M}$ is actually determined from the
low--redshift part of the Hubble diagram which should be linear
in all realistic cosmologies.
 
However, the best--fit values alone are not relevant if not supplemented
with the confidence levels for the parameters. Therefore, we performed the
estimation of model parameters using the minimisation procedure, based on
the likelihood function. We assume that supernovae measurements came with
uncorrelated Gaussian errors and in this case the likelihood function
$\mathcal{L}$ could be determined from chi--square statistic
$\mathcal{L}\propto \exp(-\chi^{2}/2)$ while probability density function
of cosmological parameters is derived from Bayes's theorem \cite{Riess98}.

Therefore, we supplement our analysis with confidence intervals in the
$(\Omega_{\lambda,0},\gamma)$ plane by calculating the marginal probability
density functions
\begin{equation}
\mathcal{P}(\Omega_{\lambda,0},\gamma)\propto \int \exp(-\chi^{2}
(\Omega_{\text{m},0},\Omega_{\lambda,0},\gamma,\mathcal{M})/2) \, d\mathcal{M}
\label{eq:108}
\end{equation}
with $\Omega_{\text{m},0}$ fixed ($\Omega_{\text{m},0}=0.05, 0.3$) and
\begin{equation}
\mathcal{P}(\Omega_{\lambda,0},\gamma)\propto \iint \exp(-\chi^{2}
(\Omega_{\text{m},0},\Omega_{\lambda,0},\gamma,\mathcal{M})/2)
\, d\Omega_{\text{m},0} \, d\mathcal{M}
\label{eq:109}
\end{equation}
without fixed $\Omega_{\text{m},0}$, respectively (a proportionality sign
equals up to the normalisation constant). In order to complete the picture
we have also derived one-dimensional probability distribution functions for
$\Omega$ obtained from joint marginalisation over $\gamma$ and
$\Omega_{\lambda,0}$. The maximum value of such a PDF informs us about the
most probable value of $\Omega$ (supported by supernovae data) within the
full class of homological models.

\section{Samples used}
 
Supernovae surveys (published data) have already five years long history.
Beginning with first published samples \cite{Perlmutter,Riess98}
other data sets have been produced either by correcting original samples for
systematics or by supplementing them with new supernovae (or both).
 
Because original Perlmutter et al. and Riess et al. samples
\cite{Perlmutter,Riess98} were completed seven years ago, presently
the newer supernovae observations are used.  Knop et al.
\cite{Knop03} have reexamined the Perlmutter sample with the host-galaxy
extinction correctly applied. They chose from the Perlmutter sample these
supernovae which were the more securely spectrally identified as type Ia and
have reasonable colour measurements. They also included eleven new high
redshift supernovae and a well known sample with low redshift supernovae.
 
We have decided to test our model using this new sample of supernovae.
They make possible to distinguish a few subsets of supernovae from this sample.
We consider two of them. The first is a subset of 58 supernovae with corrected
extinction (Knop subsample 6; hereafter K6) and the second is that of 54 low
extinction supernovae (Knop subsample 3; hereafter K3). Samples C and K3 are
similarly constructed  as containing only low extinction supernovae. The
advantage of the Knop sample is that Knop's discussion of extinction correction
was very careful and as a result his sample has extinction correctly applied.
 
Another sample was presented by Tonry et al. \cite{Tonry03} who collected a
large number of supernovae data published by different authors and added eight
new high redshift SN Ia. This sample of 230 SN Ia was re-calibrated with a
consistent zero point. Wherever possible the extinction estimates and distance
fitting were recalculated. Unfortunately, one was not able to do so for the
full sample (for details see Table~8 in Ref. \cite{Tonry03}). This sample was
further improved by Barris et al. \cite{Barris03} who added 23 high redshift
supernovae including 15 at $z \ge 0.7$ thus doubling the published record of
objects at these redshifts. Tonry et al. and Barris et al. presented the data
of redshifts and luminosity distances for their supernovae sample. Therefore,
Eqs. (\ref{eq:105}) and (\ref{eq:106}) should be modified appropriately
\cite{Williams03}
\begin{equation}
\label{eq:13a}
m-M = 5\log_{10}(D_L)_{\text{Tonry}}-5\log_{10}65 + 25
\end{equation}
and
\begin{equation}
\label{eq:13b}
{\cal M}=-5\log_{10}H_0+25.
\end{equation}
For the Hubble constant $H_0=65$ km s$^{-1}$ Mpc$^{-1}$ one gets
${\cal M}=15.935$.
 
Recently Riess et al. \cite{Riess04} significantly improved their former
group sample. They discovered 16 new type Ia Supernovae. It should be noted
that 6 of these objects have $z>1.25$ (out of total number of 7 object with
so high redshifts). Moreover, they compiled a set of previously observed SNIa
relying on large, published samples, whenever possible, to reduce systematic
errors from differences in calibrations. The full Riess sample contains 186
SNIa (``Silver'' sample). On the base of quality of the spectroscopic and
photometric record for individual Supernovae, they also selected more
restricted ``Gold'' sample of 157 Supernovae.
 
Riess et al.'s  sample has been used by many researchers as a standard
dataset. However, for the sake of comparison and illustration we analysed
also earlier Knop \cite{Knop03} sample of supernovae. This seems to be useful
because, as pointed out in the literature, studies performed on different SNIa
samples often gave different results (see for example
\cite{Godlowski04,Chou04,Dabrowski04}).

\section{Constraining equation of state from distant supernovae}
 
In order to test the model we calculate the best fit with minimum $\chi^2$
as well as we estimate the model parameters using the maximum likelihood method
\cite{Riess98}. For both statistical methods we took the parameter $n$ in
the interval $[-3.33, 2]$, $\Omega_{\text{m},0}$ in the interval $[0,1]$,
$\Omega_{\text{C},0}$ in the interval $[-1, 1]$, while an interval for
$\Omega_{\Lambda,0}$ is obtained from the equation (\ref{eq:102}).
 
We have tested the models in three different classes of models. At first we
analysed the data without any prior assumption about $\Omega_{\text{m},0}$.
In the second class we assumed that (2) $\Omega_{\text{m},0}= 0.05$
while in the last class (3) we assumed that $\Omega_{\text{m},0}= 0.3$.
 
The second class was chosen as a representative of the standard knowledge of
$\Omega_{\text{m},0}$ (baryonic plus dark matter in galactic halos
\cite{Peebles03}. In the last class we have incorporated (at the level of
$\Omega_{\text{m},0}$) the prior knowledge about baryonic content of the
Universe (as inferred from the BBN considerations). Hence this class is
representative of the models in which non matter component is responsible both
for dark matter in halos and its diffuse part (dark energy).
 
It is interesting to compare this results with these obtained for the model
with $\Lambda=0$. Such model is equivalent to Cardassian model \cite{Freese02}.
Because this model was proposed as an alternative to $\Lambda$CDM, it was
immediately verified by SNIa observations \cite{Zhu03,Sen03,Godlowski04}.
However now it is possible to perform more precise test of the models using
the new Riess sample.
 
At first we have decided to test the our model (both for vanishing and not
vanishing  cosmological constant) using Knop et al. \cite{Knop03} and Riess
et al. \cite{Riess04} samples of supernovae. In all samples we marginalise
over parameter $\mathcal{M}$. It means that the Hubble parameter is
fitted from observations, too.
 
The results of two fitting procedures performed on different samples and with
different prior assumptions concerning the cosmological models are presented
in (Table~\ref{tab:1}). The detailed results of our analysis for the flat
model with vanishing cosmological constant are summarised in Table~\ref{tab:2}.
This tables refers both to the $\chi^2$ (best fit) and results from
maximum likelihood method.  In both cases we obtained
different values of ${\cal M}$ for each analysed sample. It should be noted
that values obtained in both methods are different but differences are much
smaller for Riess et al. sample than in the case of older Knop et al. sample.
 
At first we analysed full model with not vanishing cosmological constant.
For example, with in the with the maximum likelihood method, first class
of models gives values of the parameter:
$\Omega_{\mathrm{m},0}=0.38$, $\Omega_{\mathrm{C},0}=0.64$,
$\Omega_{\Lambda,0}=0.72$, $n=-2.10$,  for sample K6, while
$\Omega_{\mathrm{m},0}=0.27$, $\Omega_{\mathrm{C},0}=0.01$,
$\Omega_{\Lambda,0}=0.75$, $n=0.00$,  for sample K3.
 
With the new Riess sample we obtain:
$\Omega_{\mathrm{m},0}=0.46$, $\Omega_{\mathrm{C},0}=0.40$,
$\Omega_{\Lambda,0}=0.12$, $n=-1.10$,  for Silver sample , while
$\Omega_{\mathrm{m},0}=0.44$, $\Omega_{\mathrm{C},0}=0.34$,
$\Omega_{\Lambda,0}=0.20$, $n=-0.90$,  for Gold Sample.
 
This result mean that for Knop sample $\Omega_{C,0}$ part play only marginal
role and the model is very close to $\Lambda$CDM model. However with the
new Riess sample we obtain that $\Omega_{m,0} \simeq 0.4$ with
$n \simeq -1$ which correspond to (hyper) phantom model.
However if we assume $\Omega_{m,0}=0.3$ from independent extragalactic
estimation \cite{Peebles03}, then $\Omega_{C,0}$ is small
and $n$ is close to zero, so  $\Lambda$CDM model is favoured.
 
On Fig. 1 we present residuals plots with respect to the Einstein-de Sitter
model for $\Lambda$CDM and our model. We observed that distant SNIa should be
brighter (in our model) than in the $\Lambda$CDM model.  What is interesting is
that the Hubble diagram for our model (both for vanishing and non vanishing
$\Lambda$) intersects the corresponding $\Lambda$CDM diagram. In
such a way, the supernovae on intermediate distant are fainter then
expected in the $\Lambda$CDM model. Please note that the Riess et al. sample
contain only very few supernovae in the intermediate distance, so this
prediction should be tested with future supernovae sample when more type Ia
supernovae measurement in the intermediate distances will be available.
 
It should be noted that knowing the best-fit values alone have not enough
scientific relevance, if confidence levels for parameter intervals are not
presented, too.  Using the minimisation procedure, based on the likelihood
method we also carry out the errors of the model parameters estimation.
On the confidence level $68.3\%$ we present parameter values for samples K6
and K3, Silver and Gold (Table~\ref{tab:3}).  The density distribution
(one dimensional PDF) for model parameters obtained by marginalisation over
remaining parameters of the model are presented in Fig. 2-5. Additionally the
on the confidence level $68.3\%$ and $95.4\%$ is marked on the figures.
 
Please note that both positive and negative values of $\Omega_{\text{C},0}$
are formally possible. It is the reason why we received on the Figs 4 and 5
bimodal distributions. Figs 6 and 7  explain this situation in more details.
These figures present the confidence levels on the plane
$(\Omega_{\text{m},0}, \Omega_{\text{C},0})$ (Fig. 6) and
$(\Omega_{\text{C},0}, n)$ (Fig. 7) minimised over remaining model
parameter.
 
Fig. 7 is more complicated than in the case of the model with
$\Omega_{\Lambda}=0$ (see Fig. 4 in Ref. \cite{Godlowski04} where the
maximum likelihood procedure suggests that $n$ should be negative and
consequently $\Omega_{\text{m},0}$ is greater than $0.3$). Now we obtain also
the possibility that $n>0$ and $\Omega_{\text{m},0}>0.3$ as a result of
presence both $\Omega_{\text{C},0}$ and $\Omega_{\Lambda}$ terms.
 
The density distribution for model parameters for the model with fixed
$\Omega_{\text{m},0}=0.3$ is presented on Figs 8-10. Those figures confirmed
that in this case the model is close to $\Lambda$CDM model.
 
For the model with vanishing cosmological constant $\Omega_{\Lambda}=0$
the errors of the model parameters estimation is presented on Table~\ref{tab:4}.
For this model, using the maximum likelihood method,
we obtained with the sample K6 that $\Omega_{\text{m},0}=0.52^{+0.09}_{-0.09}$
and $n=-3.33^{+2.00}$ on the confidence level $68.3\%$. In turn, for sample K3
we obtained that $\Omega_{\text{m},0}=0.48^{+0.08}_{-0.13}$ and
$n=-0.40^{+0.77}_{-1.24}$ on the confidence level $68.3\%$. With the new Riess
et al. sample we obtain $\Omega_{\text{m},0}=0.51^{+0.04}_{-0.05}$ and
$n=-1.60^{+0.74}_{-1.10}$ with the Silver sample, while with the Gold sample
$\Omega_{\text{m},0}=0.51^{+0.04}_{-0.05}$ and $n=-1.23^{+0.73}_{-1.23}$.
The best fit procedure also suggests that $n$ should be negative and
consequently $\Omega_{\text{m},0}$ is greater than $0.3$.
 
One can see that that result obtained with maximum likelihood method, for
$\Omega_{\text{m},0}$ and $\Omega_{\text{C},0}$ are similar for all samples,
however with new Riess et al.'s sample errors in parameter estimation
significantly decreased. Please note that with assumption that $\Lambda=0$ the
model is equivalent to the Cardassian models. We confirmed our previous results
\cite{Godlowski04} on the new Riess et al.'s sample. The observations favour
the high density universe with $n$ negative. On the other hand, if we assume
$\Omega_{\text{m},0}=0.3$ then $n \simeq 0$ which correspond to $\Lambda$CDM
model, like for scaling multifluids model with non vanishing $\Lambda$.
 
For the  model with $\Omega_{\text{m},0}=0.3$ we obtained for sample K6
$n = -0.13$ with $\sigma(n)=0.23$. In turn, for sample K3 we obtained
$n =-0.20$, $\sigma(n)=0.17$. With the new Riess et al. sample we obtained
$n = -0.07$ with $\sigma(n)=0.10$ for Silver sample, while $n =-0.03$,
$\sigma(n)=0.10$ for Gold sample.
On can see that result obtained with both samples are similar but with the
Riess et al. sample errors in estimation of the parameters significantly
decreased. These results mean that if we assume $\Omega_{\text{m},0}=0.3$
then $n \simeq 0$ which correspond to $\Lambda$CDM model.

In this way, it is crucial to determine which combination of parameters
give the preferred fit to data. This is the statistical problem of
model selection \cite{Liddle04}. The problem is usually the elimination of
parameters which play insufficient role in improving the fit data available.
Important role in this area plays especially the Akaike
information criterion AIC \cite{Akaike74}. This criterion is defined as
\begin{equation}
\label{eq:244}
AIC=-2\ln{\mathcal{L}}+2k
\end{equation}
where $\mathcal{L}$ is the maximum likelihood and $k$ is the number of the
parameter of the model. The best model is the model which minimises the AIC.
 
The Bayesian information criterion BIC introduced by Schwarz
\cite{Schwarz:1978}  is defined as:
\begin{equation}
\label{eq:12}
\mathrm{BIC} = - 2\ln{\mathcal{L}} + d\ln{N}
\end{equation}
where $N$ is the number of data points used in the fit. While AIC tends to
favour models with large number of parameters, the BIC the more strongly
penalises them, so BIC provides the useful approximation to full evidence
in the case of no prior on the set of model parameters \cite{Parkinson:2005}.

The effectiveness of using these criteria in the current cosmological
applications has been recently demonstrated by Liddle \cite{Liddle04}
Please note that both information criteria values have no absolute sense and
only the relative values between different models are physically interesting.
For the BIC a difference of $2$ is treated as a positive evidence
($6$ as a strong evidence) against the model with larger value of BIC
\cite{Jeffreys:1961,Mukherjee:1998wp}.

 The AIC for the models under consideration is presented in the
(Table~\ref{tab:5}) while for BIC in the (Table~\ref{tab:6}).
The BIC information criterion do not show significant differences between
scaling multifluid model with vanishing $\Lambda$ and $\Lambda$CDM model.
The model which minimising AIC is scaling multifluids
model with the vanishing cosmological constant (equivalent to the
Cardassian model). It means that, from the statistical point of view,
if we compare scaling multifluid models with vanishing and non-vanishing
$\Lambda$, the extra term with $\Lambda$ does not improve significantly
the quality of the fit.
 
Basing on AIC information criterion,
it is clear that the scaling multifluid model with vanishing
$\Lambda$ fits better to data then the $\Lambda$CDM model.
However scaling multifluid model with vanishing $\Lambda$ indicate
high density universe, with $\Omega_{m,0} \simeq 0.5$ which
seems to be too high with comparison with present extragalactic data
\cite{Peebles03}. Scaling multifluid model with non-vanishing $\Lambda$
allow density of the universe $\Omega_{m,0}$ close to 0.3 (Fig.7)
while $\Lambda$CDM model predict $\Omega_{m,0} \simeq 0.3$.
It clearly shows that more precise measurements of $\Omega_{m,0}$
from independent observations is necessary to final discriminate
between the $\Lambda$CDM and scaling multifluid models.

\section{Conclusion}
 
The Supernova Cosmology Project and the High--Z--Supernova Search reported of
their observations of type Ia supernovae and suggest that the expansion of
the Universe is still accelerating due to the presence of unknown form of
matter called dark energy. For the accelerating Universe the equation of state
parameter $w=\frac{p}{\rho}$ for dark must satisfy $w<-\frac{1}{3}$. The
cosmological constant $\Lambda$ is arguable, but at the some time the simplest
candidate for this dark energy, although it is well known, that predictions
for its value are many orders of magnitude off from the observationally acceptable
value. The introducing of $\Lambda=\Lambda(a)$ or quintessence is usually
proposed as a possible solution to avoid the cosmological constant problem.
 
In presented paper the idea of derivation the form of equation of state from
symmetries of self-similarity of the FRW dynamics is considered. We have shown
that these symmetries enforce appropriate equation of state used commonly in
cosmology. The property of homology is very important in astrophysics, when main
sequence in the Hertzsprung--Russell diagram can be reconstructed from the
invariants of homology transformations. Moreover from the Str\"omgren's theorem,
new solution of stellar structure equation can be obtained from the known
ones through the homologous transformation. The new solutions describe new
configurations with different masses, radius and chemical compositions
(the so--called homologous stars). For recently obtained results see
\cite{Stromgren37,Szydlowski04} where this problem is addressed in the context
of brane cosmology. We pointed out that our Universe is also homological
provided that it is filled by matter with the form of equation of state for
noninteracting mixture of fluids $p=-\Lambda+C_{1}\rho(a)+C_{2}a^{\delta}$ and
$\rho=\Lambda+\rho_{01}a^{-3(1+w)}+\rho_{02}a^{\delta}+\rho_{03}a^{-3}$.
This form is commonly used in cosmological considerations. The key idea of
this work is derivation this form of the equation of state from the first
principles. It is proposed to derivate $p=p(\rho)$ from postulate of
self-similarity of its dynamics. This property is well known to engineers who
build ship models as a prototype of real ship. It is justified by the fact
that Navier-Stoke's equation are invariant against similarity symmetries.
 
As a result we obtain the commonly used form of equation of state for mixture
of noninteracting, dust matter, dark energy (the cosmological constant) and
cosmic time variation of cosmological constant parameterised by the scale factor.
Therefore subliminal role of mathematics can also be seen in cosmology when
we are looking for adequate form of equation of state for dark energy.

We showed that the model with scaling multifluids fits well the supernovae data.
For simplicity of presentation we demonstrate this for the case
$C_1=\rho_{01}=0$. Physically it means that the Universe is filled generally by
dust matter, cosmological constant fluid and additional scaling fluid which
comes from Cardassian modification of the FRW equation (or equivalently it
it is scaling fluid describing phantom fields for example).
 
For the scaling multifluids model with $\Lambda \ne 0$,  we found that
$\Omega_{\text{m},0} \simeq 0.4$ with $n \simeq -1$ which correspond to
(hyper) phantom model. If we assume $\Omega_{\text{m},0}=0.3$ from independent
extragalactic estimation, then $\Omega_{\text{C},0}$ is small while value of
$n$ close to zero is favoured and model becomes close to $\Lambda$CDM.
With assumption that $\Lambda=0$ the model is equivalent to Cardassian
models. We confirmed our previous results \cite{Godlowski04} on the new Riess
sample. In particular the observations favour high density Universe with $n$
negative. Again, if we assume $\Omega_{\text{m},0}=0.3$ then $n \simeq 0$
which correspond to $\Lambda$CDM model.
 
>From Fig. 1 can be seen the $m$-$z$ relation for our model and
$\Lambda$CDM one. We observed that distant SNIa should be brighter
(in our model) than in the $\Lambda$CDM model. What is interesting that the
Hubble diagram for the model under consideration intersects the corresponding
$\Lambda$CDM diagram. In such a way the supernovae on intermediate distant are
fainter then expected in $\Lambda$CDM model. This predictions could be tested
with the future supernovae data.
 
Our results demonstrate the existence of alternative model to $\Lambda$CDM
model in explanation SNIa data. Therefore it should be interesting to compare
both models from the point of view Akaike information criterion. Our result
show that scaling multifluids model with vanishing $\Lambda$ significantly
better fits data then $\Lambda$CDM model.
 
Moreover the cosmological model filled by scaling fluid makes a step toward
solving the coincidence problem of the present value of dark matter and
dark energy components \cite{Amendola00}.
 
To make the ultimate decision which model describes our Universe it is
necessary to obtain the precise value of $\Omega_{m,0}$ from independent
observations because $\Lambda$CDM model and scaling multifluid models
predict different density of the Universe.

\section{Acknowledgements}
 
M.S. was supported by KBN grant 1 PO3D 003 26.

\begin{table}
\noindent
\caption{Results of the statistical analysis of the model obtained both for
the Knop et al. and Riess et al. samples from the best fit with minimum
$\chi^2$ (denoted with BF) and from the likelihood method (denoted with L).
The same analysis was repeated with fixed $\Omega_{\text{m},0}$.}
\label{tab:1}
\begin{tabular}{@{}p{1.5cm}rrrrrrr}
\hline \hline
sample & $\Omega_{\mathrm{m},0}$ & $\Omega_{\mathrm{C},0}$&  $n$&
$\Omega_{\Lambda,0}$ &  $\mathcal{M}$ & $\chi^2$& method \\
\hline
 K6   &  0.64 & 0.58 &-3.17 &-0.22 &-3.61 & 53.4 &  BF  \\
      &  0.38 & 0.64 &-2.10 & 0.72 &-3.53 & ---  &  L   \\
      &  0.05 & 0.03 & 2.60 & 0.92 &-3.53 & 55.1 &  BF  \\
      &  0.05 & 0.03 & 0.43 & 0.92 &-3.51 & ---  &  L   \\
      &  0.30 & 0.13 &-3.33 & 0.57 &-3.55 & 55.1 &  BF  \\
      &  0.30 & 0.00 & 0.00 & 0.70 &-3.52 & ---  &  L   \\
\hline
 K3   &  0.45 & 0.98 &-0.47 &-0.43 &-3.49 & 60.3 &  BF  \\
      &  0.27 & 0.01 & 0.00 & 0.75 &-3.47 & ---  &  L   \\
      &  0.05 & 0.05 & 1.97 & 0.90 &-3.48 & 60.4 &  BF  \\
      &  0.05 & 0.03 & 0.40 & 0.92 &-3.46 & ---  &  L   \\
      &  0.30 & 0.13 &-2.23 & 0.57 &-3.49 & 60.5 &  BF  \\
      &  0.30 & 0.00 & 0.00 & 0.69 &-3.46 & ---  &  L   \\
\hline
Silver&  0.44 & 0.32 &-3.27 & 0.24 &15.895&226.7 &  BF  \\
      &  0.46 & 0.40 &-1.10 & 0.12 &15.915& ---  &  L   \\
      &  0.05 & 0.07 & 1.77 & 0.88 &15.935&229.4 &  BF  \\
      &  0.05 & 0.07 & 1.50 & 0.88 &15.945& ---  &  L   \\
      &  0.30 & 0.10 &-3.33 & 0.60 &15.915&230.9 &  BF  \\
      &  0.30 & 0.09 & 0.00 & 0.61 &15.945& ---  &  L   \\
\hline
Gold  &  0.43 & 0.28 &-3.33 & 0.29 &15.905&172.1 &  BF  \\
      &  0.44 & 0.34 &-0.90 & 0.20 &15.935& ---  &  L   \\
      &  0.05 & 0.10 & 1.57 & 0.85 &15.945&174.0 &  BF  \\
      &  0.05 & 0.09 & 1.30 & 0.86 &15.945& ---  &  L   \\
      &  0.30 & 0.07 &-3.33 & 0.63 &15.925&175.2 &  BF  \\
      &  0.30 & 0.00 & 0.00 & 0.70 &15.945& ---  &  L   \\
\hline
\end{tabular}
\end{table}

\begin{table}
\noindent
\caption{Results of the statistical analysis of the model with fixed
$\Omega_{\Lambda,0}=0$ obtained both for the Knop et al. and Riess et al.
samples from the best fit with minimum $\chi^2$ (denoted with BF) and from
the likelihood method (denoted with $L$). The same analysis was repeated
with fixed $\Omega_{\text{m},0}$.}
\label{tab:2}
\begin{tabular}{@{}p{1.5cm}rrrrrrr}
\hline \hline
sample & $\Omega_{\mathrm{m},0}$ & $\Omega_{\mathrm{C},0}$&  $n$&
$\Omega_{\Lambda,0}$ &  $\mathcal{M}$ & $\chi^2$& method \\
\hline
 K6   &  0.54 & 0.46 &-3.33 & 0.   &-3.60 & 53.5 &  BF  \\
      &  0.52 & 0.48 &-3.33 & 0.   &-3.55 & ---  &  L   \\
      &  0.05 & 0.95 & 0.33 & 0.   &-3.51 & 56.3 &  BF  \\
      &  0.05 & 0.95 & 0.30 & 0.   &-3.51 & ---  &  L   \\
      &  0.30 & 0.70 &-0.10 & 0.   &-3.52 & 55.6 &  BF  \\
      &  0.30 & 0.70 &-0.13 & 0.   &-3.53 & ---  &  L   \\
\hline
 K3   &  0.42 & 0.58 &-0.77 & 0.   &-3.49 & 60.3 &  BF  \\
      &  0.48 & 0.52 &-0.40 & 0.   &-3.49 & ---  &  L   \\
      &  0.05 & 0.95 & 0.30 & 0.   &-3.46 & 61.5 &  BF  \\
      &  0.05 & 0.95 & 0.30 & 0.   &-3.46 & ---  &  L   \\
      &  0.30 & 0.70 &-0.13 & 0.   &-3.47 & 60.6 &  BF  \\
      &  0.30 & 0.70 &-0.17 & 0.   &-3.48 & ---  &  L   \\
\hline
Silver&  0.50 & 0.50 &-1.73 & 0.   &15.905&227.1 &  BF  \\
      &  0.51 & 0.49 &-1.60 & 0.   &15.905& ---  &  L   \\
      &  0.05 & 0.95 & 0.40 & 0.   &15.975&239.3 &  BF  \\
      &  0.05 & 0.95 & 0.40 & 0.   &15.965& ---  &  L   \\
      &  0.30 & 0.70 &-0.07 & 0.   &15.945&232.3 &  BF  \\
      &  0.30 & 0.70 &-0.07 & 0.   &15.945& ---  &  L   \\
\hline
Gold  &  0.49 & 0.51 &-1.37 & 0.   &15.915&172.5 &  BF  \\
      &  0.51 & 0.49 &-1.23 & 0.   &15.915& ---  &  L   \\
      &  0.05 & 0.95 & 0.43 & 0.   &15.975&180.8 &  BF  \\
      &  0.05 & 0.95 & 0.43 & 0.   &15.975& ---  &  L   \\
      &  0.30 & 0.70 &-0.03 & 0.   &15.945&175.9 &  BF  \\
      &  0.30 & 0.70 &-0.03 & 0.   &15.945& ---  &  L   \\
\hline
\end{tabular}
\end{table}
 
\begin{table}
\caption{Model parameter values obtained from the minimisation procedure
carried out with the Knop et al. and Riess et al. samples.}
\label{tab:3}
\begin{tabular}{@{}p{1.5cm}rrrr}
\hline \hline
sample & $\Omega_{\mathrm{m},0}$& $\Omega_{\mathrm{C},0}$ &
 $\Omega_{\Lambda,0}$ & $n$ \\
\hline
 K6   & $0.38^{+0.32}_{-0.18}$ & $0.64^{+0.36}_{-0.48}$
      & $0.72^{+0.30}_{-0.32}$ & $-2.10^{+1.63}_{-1.23}$ \\
      & $ 0.05 $ & $0.03^{+0.38}_{-0.76}$
      & $0.92^{+0.76}_{-0.38}$ & $ 0.43^{+0.97}_{-2.00}$ \\
      & $ 0.3 $ & $0.00^{+0.38}_{-0.23}$
      & $0.70^{+0.23}_{-0.38}$ & $ 0.00^{+0.58}_{-1.70}$ \\
 K3   & $0.27^{+0.20}_{-0.13}$ & $0.01^{+0.63}_{-0.38}$
      & $0.75^{+0.38}_{-0.70}$ & $ 0.00^{+0.93}_{-1.56}$ \\
      & $0.05 $ & $0.03^{+0.38}_{-0.76}$
      & $0.92^{+0.76}_{-0.38}$ & $ 0.40^{+0.96}_{-1.90}$ \\
      & $0.3 $ & $0.00^{+0.38}_{-0.20}$
      & $0.69^{+0.21}_{-0.37}$ & $ 0.00^{+0.50}_{-1.60}$ \\
Silver& $0.46^{+0.10}_{-0.08}$ & $0.40^{+0.34}_{-0.22}$
      & $0.12^{+0.30}_{-0.44}$ & $-1.10^{+0.50}_{-1.75}$ \\
      & $0.05 $ & $0.07^{+0.09}_{-0.04}$
      & $0.88^{+0.04}_{-0.09}$ & $ 1.50^{+0.40}_{-0.47}$ \\
      & $0.3 $ & $0.09^{+0.17}_{-0.20}$
      & $0.61^{+0.20}_{-0.17}$ & $ 0.00^{+0.36}_{-1.20}$ \\
Gold  & $0.44^{+0.11}_{-0.11}$ & $0.34^{+0.39}_{-0.33}$
      & $0.20^{+0.55}_{-0.47}$ & $-0.90^{+0.60}_{-1.90}$ \\
      & $0.05 $ & $0.09^{+0.15}_{-0.05}$
      & $0.86^{+0.05}_{-0.15}$ & $ 1.30^{+0.43}_{-0.47}$ \\
      & $0.3 $ & $0.00^{+0.26}_{-0.17}$
      & $0.70^{+0.17}_{-0.26}$ & $ 0.00^{+0.50}_{-1.60}$ \\
\end{tabular}
\end{table}
 
\begin{table}
\caption{Model parameter values obtained from the minimisation procedure
carried out on the Knop and Riess samples. Model with fixed
$\Omega_{\Lambda,0}=0$.}
\label{tab:4}
\begin{tabular}{@{}p{1.5cm}rrrr}
\hline \hline
sample & $\Omega_{\mathrm{m},0}$& $\Omega_{\mathrm{C},0}$ &
 $\Omega_{\Lambda,0}$ & $n$ \\
\hline
 K6   &  $ 0.52^{+0.09}_{-0.09}$ & $0.48^{+0.09}_{-0.09}$
      & $0.$ & $-3.33^{+2.00}_{-0.00}$ \\
 K3   & $0.48^{+0.08}_{-0.13}$ & $0.52^{+0.13}_{-0.08}$
      & $0.$ & $-0.40^{+0.77}_{-1.24}$ \\
Silver& $ 0.51^{+0.04}_{-0.05}$ & $0.49^{+0.05}_{-0.04}$
      & $0.$ & $-1.60^{+0.74}_{-1.10}$ \\
Gold  & $0.51^{+0.04}_{-0.05}$ & $0.49^{+0.05}_{-0.04}$
      & $0.$ & $-1.23^{+0.73}_{-1.23}$ \\
\end{tabular}
\end{table}

\begin{table}
\noindent
\caption{The Akaike information criterion (AIC) for models:
$\Lambda$CDM model ($\Lambda$CDM), Cardassian and Scaling Multifluids.}
\label{tab:5}
\begin{tabular}{cccc}
\hline \hline
sample & $\Lambda$CDM & Cardassian & Scaling Multifluids \\
\hline
K6              & 59.8   &  59.5 &  61.4  \\
K3              & 64.3   &  66.3 &  68.3  \\
Silver          & 236.6  & 233.1 & 234.7  \\
Gold            & 179.9  & 178.5 & 180.1  \\
\hline
\end{tabular}
\end{table}

\begin{table}
\noindent
\caption{The Bayesian information criterion (BIC) for models:
$\Lambda$CDM model ($\Lambda$CDM), Cardassian and Scaling Multifluids.}
\label{tab:6}
\begin{tabular}{cccc}
\hline \hline
sample & $\Lambda$CDM & Cardassian & Scaling Multifluids \\
\hline
K6              & 63.8   &  65.5 &  69.4  \\
K3              & 68.3   &  72.3 &  76.3  \\
Silver          & 243.0  & 242.8 & 247.6  \\
Gold            & 186.0  & 186.7 & 196.3  \\
\hline
\end{tabular}
\end{table}
 
\begin{figure}
\includegraphics[width=0.8\textwidth]{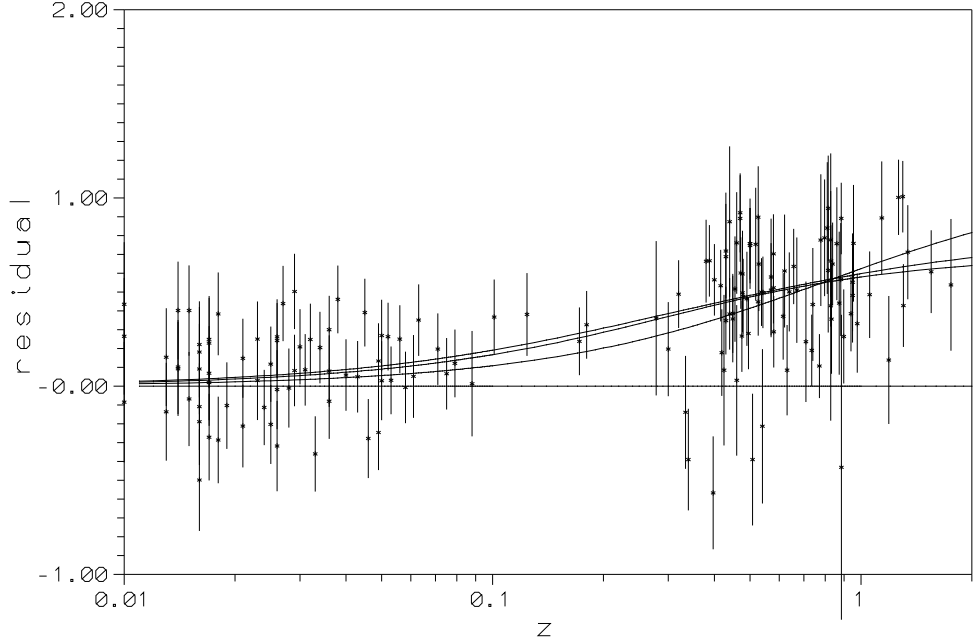}
\caption{Riess Gold sample, {\cal M}=15.935
Residuals (in mag) between the Einstein-de Sitter model and four
cases: the Einstein-de Sitter itself (zero line), the $\Lambda$CDM flat model
(upper curve), the best-fitted  model (upper-middle curve),
$\Omega_{\mathrm{m},0}=0.43$, $\Omega_{C,0}=0.28$, $\Omega_{\Lambda,0}=0.29$,
$n=-3.33$, and the best-fitted $\Lambda=0$ model (lower-middle curve)
$\Omega_{\mathrm{m},0}=0.49$, $\Omega_{C,0}=0.51$, $n=-1.37$.}
\label{fig:1}
\end{figure}

\begin{figure}
\includegraphics[width=0.6\textwidth]{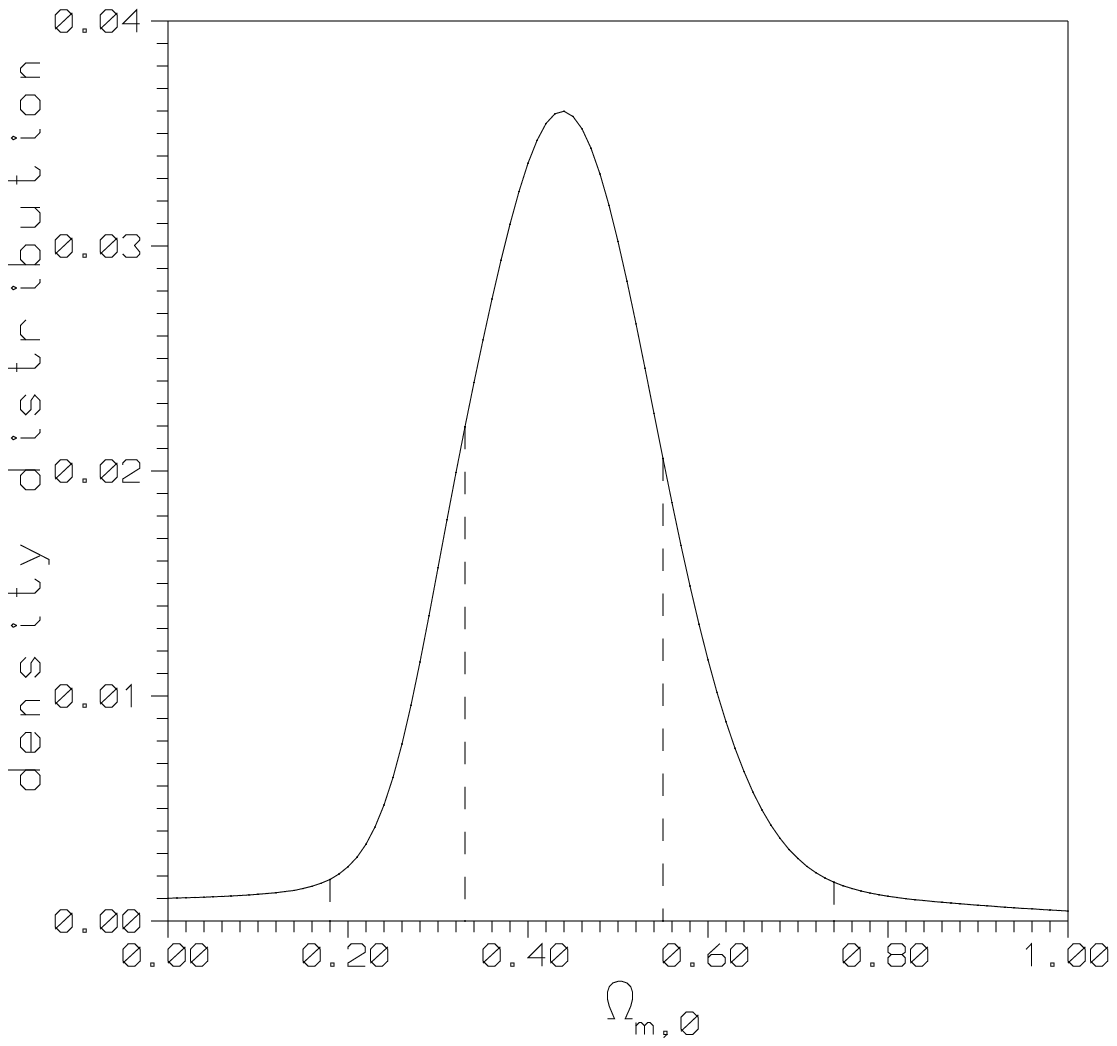}
\caption{The density distribution for
$\Omega_{\mathrm{m},0}$.
We obtain that $\Omega_{\mathrm{m},0}=0.44^{+0.11}_{-0.11}$
on the confidence level $68.3\%$ (the inner dash lines). Additionally the
confidence level $95.4\%$ is marked (the outer dash lines).}
\label{fig:2}
\end{figure}

\begin{figure}
\includegraphics[width=0.6\textwidth]{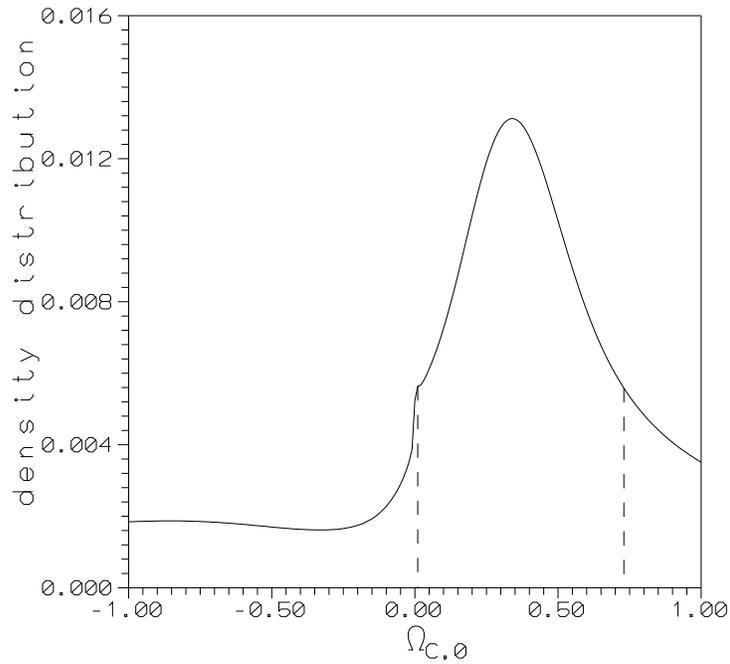}
\caption{The density distribution for
$\Omega_{C,0}$.
We obtain that $\Omega_{C,0}=0.34^{+0.39}_{-0.33}$
on the confidence level $68.3\%$ (the inner dash lines).
Both positive and negative values of $\Omega_{C,0}$ are formally possible.}
\label{fig:3}
\end{figure}

\begin{figure}
\includegraphics[width=0.6\textwidth]{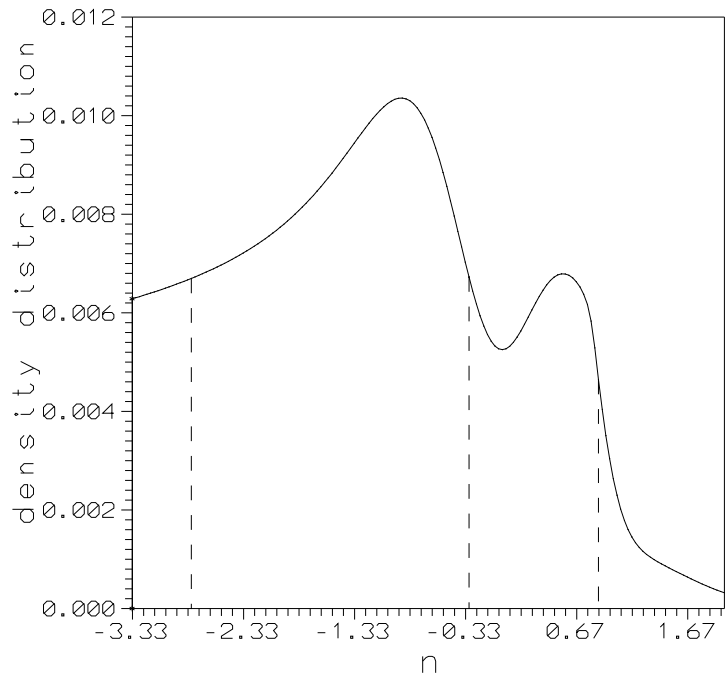}
\caption{The density distribution for $n$.
We obtain that $n=-0.90^{+0.60}_{-1.90}$ on the confidence level $68.3\%$ (the
inner dash lines). Additionally the confidence level $95.4\%$ is marked (the
outer dash lines). Both positive and negative values of $n$ are formally
possible.}
\label{fig:4}
\end{figure}

\begin{figure}
\includegraphics[width=0.6\textwidth]{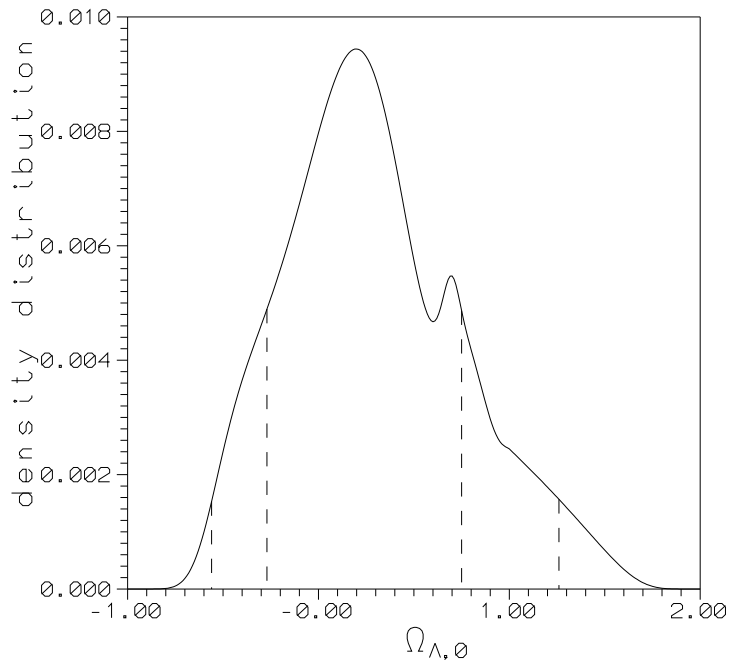}
\caption{The density distribution for
$\Omega_{\Lambda,0}$.
We obtain that $\Omega_{\Lambda,0}=0.20^{+0.55}_{-0.47}$
on the confidence level $68.3\%$ (the inner dash lines). Additionally the
confidence level $95.4\%$ is marked (the outer dash lines).}
\label{fig:5}
\end{figure}

\begin{figure}
\includegraphics[width=0.6\textwidth]{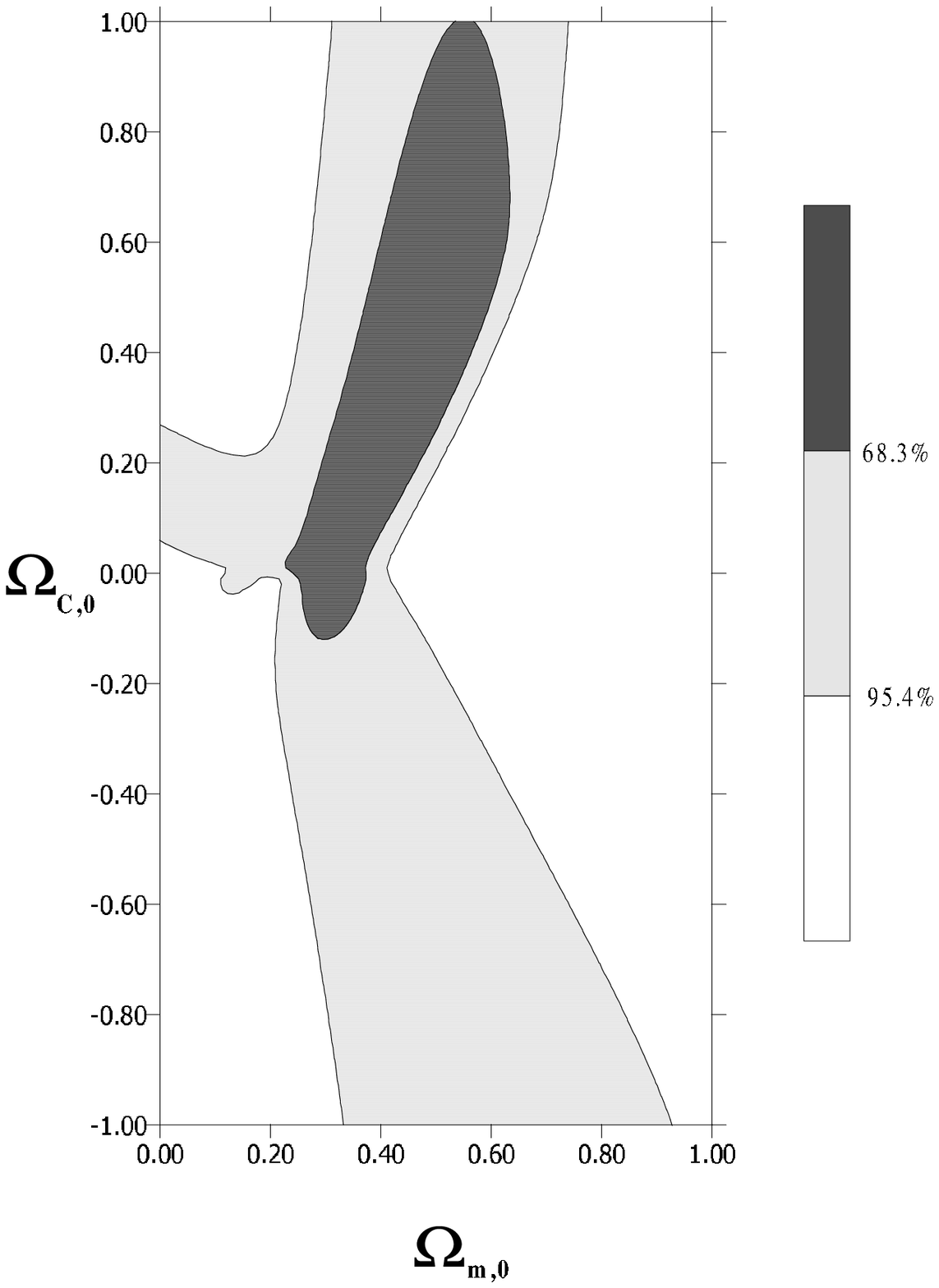}
\caption{Confidence levels on the plane
$(\Omega_{\mathrm{m},0}, \Omega_{\mathrm{C},0})$ minimised over remaining
model parameter.
The figure shows of the preferred value of
$\Omega_{\mathrm{m},0}$ and $\Omega_{\mathrm{C},0}$.}
\label{fig:6}
\end{figure}

\begin{figure}
\includegraphics[width=0.6\textwidth]{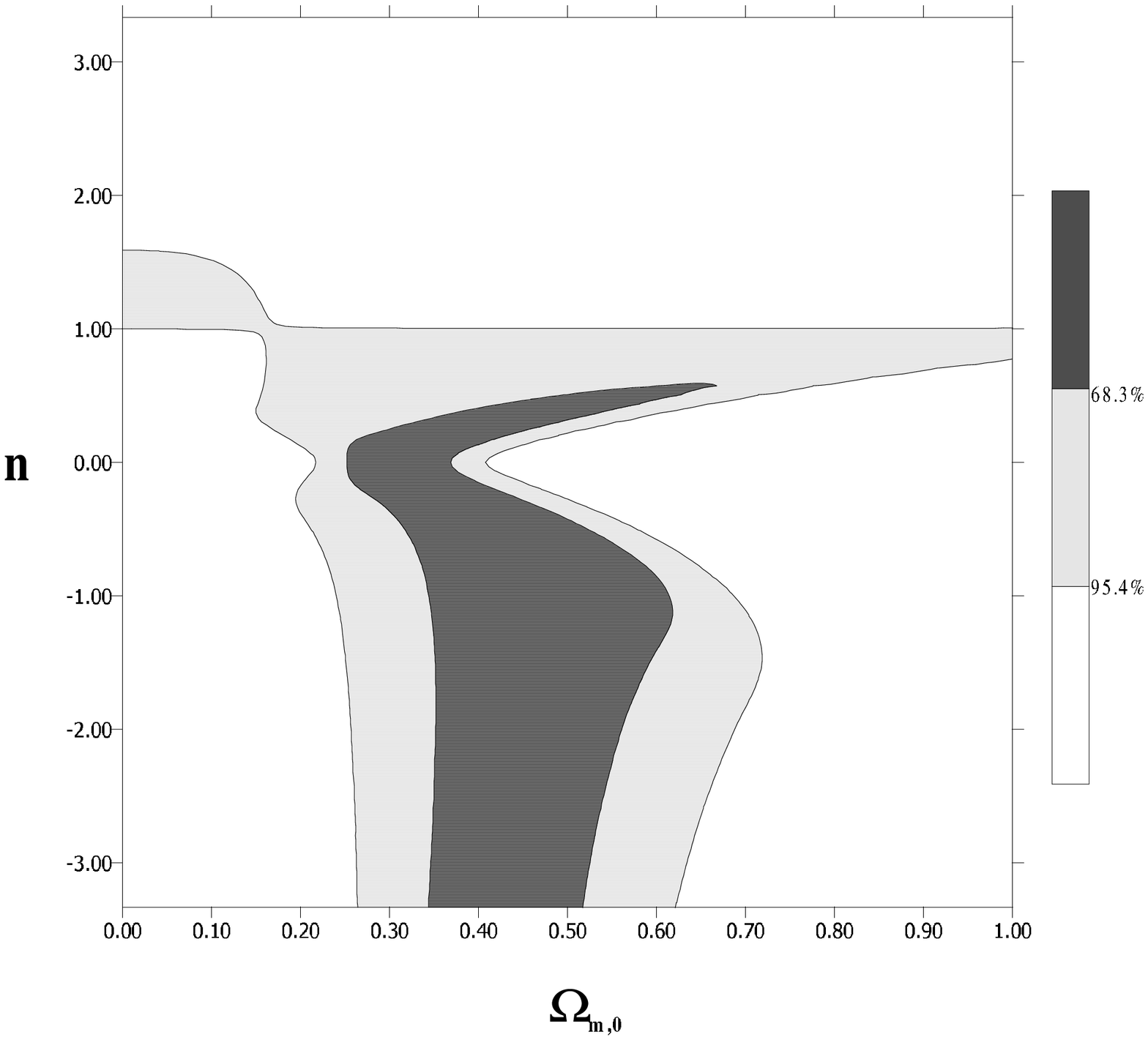}
\caption{Confidence levels on the plane
$(\Omega_{\mathrm{m},0}, n)$ minimised over remaining
model parameter.
The figure shows of the preferred value of
$\Omega_{\mathrm{m},0}$ and $n$.}
\label{fig:7}
\end{figure}

\begin{figure}
\includegraphics[width=0.6\textwidth]{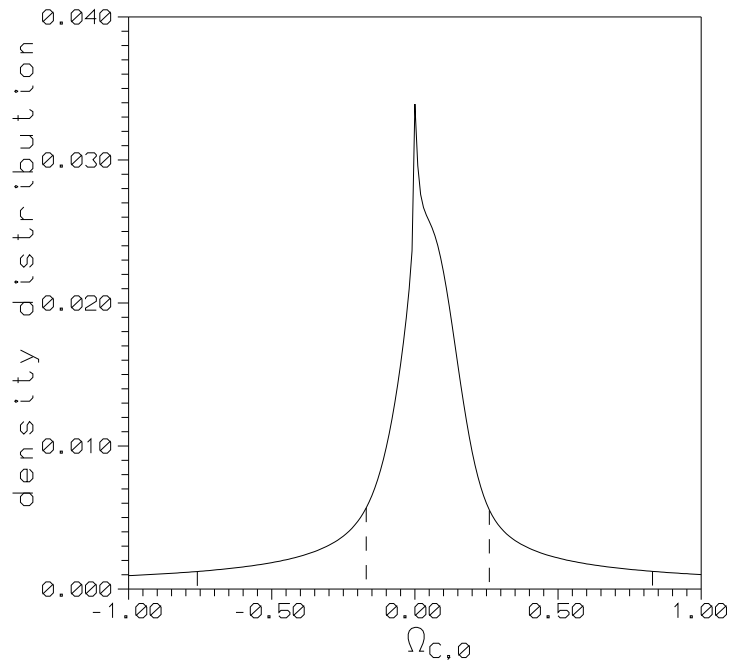}
\caption{The density distribution for
$\Omega_{C,0}$
for the model with $\Omega_{\mathrm{m},0}$=0.3.
We obtain that $\Omega_{C,0}=0.00^{+0.26}_{-0.17}$
on the confidence level $68.3\%$ (the inner dash lines). Additionally the
confidence level $95.4\%$ is marked (the outer dash lines).
Both positive and negative values of $\Omega_{C,0}$ are formally possible.}
\label{fig:8}
\end{figure}

\begin{figure}
\includegraphics[width=0.6\textwidth]{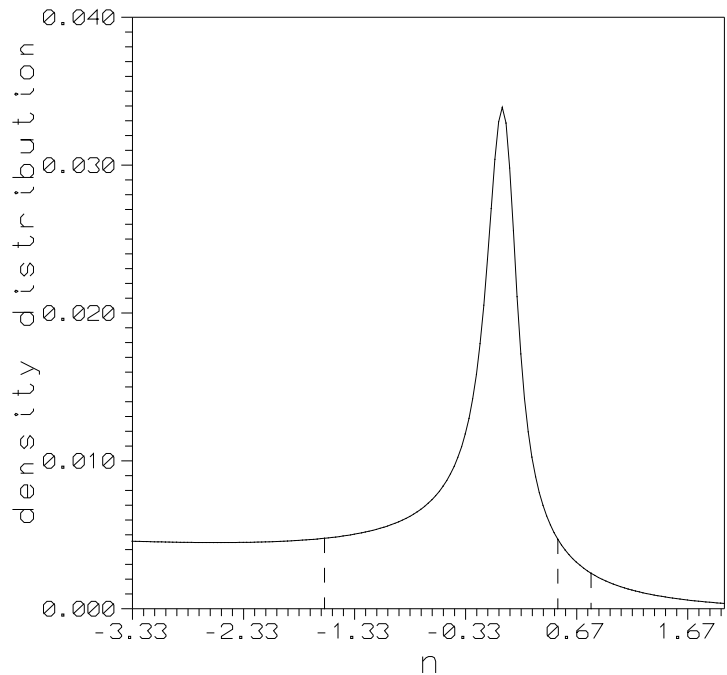}
\caption{The density distribution for $n$
for the model with $\Omega_{\mathrm{m},0}$=0.3.
We obtain that $n=0.00^{+0.50}_{-1.60}$ on the confidence level $68.3\%$ (the
inner dash lines). Additionally the confidence level $95.4\%$ is marked (the
outer dash lines).}
\label{fig:9}
\end{figure}

\begin{figure}
\includegraphics[width=0.6\textwidth]{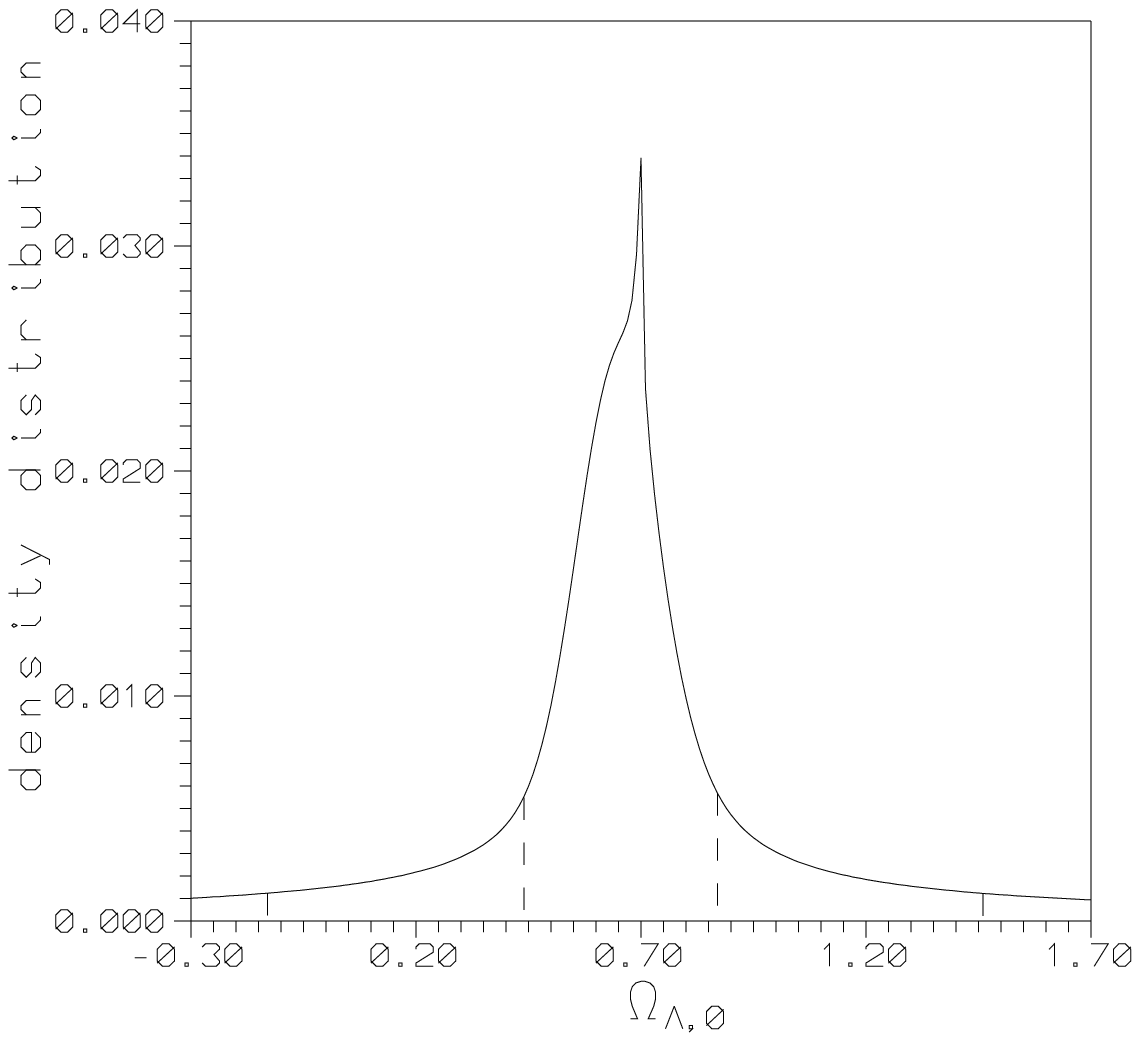}
\caption{The density distribution for
$\Omega_{\Lambda,0}$
for the model with $\Omega_{\mathrm{m},0}$=0.3.
We obtain that $\Omega_{\Lambda,0}=0.70^{+0.17}_{-0.26}$
on the confidence level $68.3\%$ (the inner dash lines). Additionally the
confidence level $95.4\%$ is marked (the outer dash lines).}
\label{fig:10}
\end{figure}


\begin{thebibliography}{99}
 
\bibitem{Sachs77}
Sachs R.~K., Wu H. {\sl Gen. Relat. for Math.} 1977
 
\bibitem{Heller83}
Szydlowski M., Heller M., {\sl Acta Phys. Pol}~{\bf B14} 571 1983
 
\bibitem{McCrea51}
McCrea,  W. H., {\sl Proc. R.Soc. London}~{\bf A206}, 569, 1951
 
\bibitem{Peebles03}
Peebles P.~J.~E., Ratra B., {\sl Rev. Mod. Phys.}{\bf 75}, 559, 2003
 
\bibitem{Perlmutter}
Perlmutter S., Aldering G., Goldhaber G., et al., {\sl Astrophys. J}~{\bf 517},
565, 1999
 
\bibitem{Riess98}
Riess A., et al., {\sl Astron. J}~{\bf 116}, 1009, 1998
 
\bibitem{Stephani89}
Stephani H.,  {\sl Differential equation - Their solution using
symmetries}, eds MacCallum M., Cambridge University Press, Cambridge, 1989;
 
\bibitem{Aug03a}
Aguirregabiria J. M., et al. {\sl Phys. Rev.}~{\bf D67}, 083518, 2003
 
\bibitem{Aug03b}
Chimento L. P., {\sl Phys. Rev.}~{\bf D65}, 0633517, 2002
 
\bibitem{Belinchon00}
Belinchon J. A., Davila, {\sl Class. Quantum Grav.}~{\bf 17}, 3183, 2000
 
\bibitem{Belinchon02}
Belinchon J. A., Harko T., Mak M.K. {\sl Class. Quantum Grav.}~{\bf 19}, 3003, 2002
 
\bibitem{Arnold80}
Arnold V. {\sl Chaptiers supplementataires de la theorie des {\'e}quations
differentiales ordinaries}, Mir, Moscow, 1980.
 
\bibitem{Collins77a}
Collins C. B. {\sl Gen. Relativ. Grav.}~{\bf 8}, 717, 1977
 
\bibitem{Collins77b}
Collins C. B. {\sl J. Math. Phys.}~{\bf 18}, 1374, 1977
 
\bibitem{Aug03c}
van den Hoogen R.J., Coley A.A., Wands D. {\sl Class. Quantum Grav.}~{\bf 16}, 1843, 1999
 
\bibitem{Carr99}
Carr B.J., Coley A.A. {\sl Class. Quantum Grav.}~{\bf 16}, R31, 1999
 
\bibitem{Maeda04}
Maeda H., Harada T. gr-qc/0405113
 
\bibitem{Carr93}
Carr B.J., unpublished 1993
 
\bibitem{Harada01}
Harada T., Maeda H. {\sl Phys. Rev.}~{\bf D63}, 084022, 2001
 
\bibitem{Choptuik93}
Choptuik M.W. {\sl Phys. Rev. Lett.}~{\bf 70}, 9, 1993
 
\bibitem{Guo04}
Guo Z.K., Zhang Y.Z. astro-ph/0411524
 
\bibitem{Tsujikawa04}
Tsujikawa S., Sami M., hep-th/0409212
 
\bibitem{Gumjudpai05}
Gumjudpai B., Naskar T., Sami M., Tsujikawa S. hep-th/0502191
 
\bibitem{Amendola00}
Amendola L., Tocchini- Valentini D. astro-ph/0011243
 
\bibitem{Sahni02}
Sahni V. {\sl Class. Quantum Grav.}~{\bf 19} 3435, 2002
 
\bibitem{Amendola99}
Amendola L., astro-ph/9904120
 
\bibitem{Amendola04}
Amendola L., Gasparini M., Piazza F. astro-ph/0407573
 
\bibitem{Majerrotto04}
Majerrotto E., Sapone D., Amendola L. astro-ph/0410543
 
\bibitem{Ellis05}
Ellis G.F.R., Buchert T., gr-qc/0506106
 
\bibitem{Harko04}
Harko T., Mak M. K. {\sl Phys. Rev.}~{\bf D69}, 064020, 2004
 
\bibitem{Copeland98}
Copeland E.J., Liddle A.R., Wands D. {\sl Phys. Rev.}~{\bf D57}, 4686, 1998
 
\bibitem{Falle91}
Falle S. {\sl Mon. Not. R. Astr. Soc.}~{\bf 250}, 581 1991),
 
\bibitem{Ibragimov83}
Ibragimov N. K. {\sl Transformations in Mathematical Physics} Moscow 1983
 
\bibitem{Hydon99}
Hydon P. E. {\sl Symmetry Methods for Differential Equation},
Cambridge University Press, Cambridge, 1999
 
\bibitem{Biesiada89}
Biesiada M., Szydlowski M., Szczesny J. {\sl Acta Cosmologica}~{\bf XVI},
115, 1989
 
\bibitem{starkman}
Lue A., Starkman G. D., {\sl Phys. Rev. Lett.}~{\bf 92}, 131102, 2004
 
\bibitem{Jafarizadek99}
Jafarizadeh M. A., et al. {\sl Phys. Rev.}~{\bf D60}, 063514, 1999
 
\bibitem{Silveira97}
Silveira V., Waga I. {\sl Phys. Rev.}~{\bf D56}, 4625, 1997
 
\bibitem{Alam03}
Alam U., Sahni V., Starobinsky A. A. {\sl JCAP}~{\bf 04}, 002, 2003
 
\bibitem{Gorini04}
Gorini V., Kamenschik A., Morchella U., Pasquier V., gr-qc/0403062, 2004
 
\bibitem{Godlowski03}
Godlowski W., Szydlowski M., {\sl Gen. Relat. Grav.}~{\bf 35}, 2171, 2003
 
\bibitem{Godlowski04}
Godlowski W., Szydlowski M., Krawiec A., {\sl Astrophys. J}~{\bf 605},
599, 2004
 
\bibitem{Godlowski04a}
Godlowski W., Szydlowski M., {\sl Gen. Relat. Grav.}~{\bf 36}, 767, 2004
 
\bibitem{Dabrowski04}
Dabrowski M.P., Godlowski W., Szydlowski M. {\sl Int. J. Mod. Phys.}~{\bf D13},
1669, 2004
 
\bibitem{Godlowski04b}
Godlowski W., Stelmach J., Szydlowski M. {\sl Class. Quantum Grav.},
{\bf 21}, 3953, 2004
 
\bibitem{Biesiada05}
Biesiada M., Godlowski W., Szydlowski M., {\sl Astrophys. J.}~{\bf 622} 28, 2005
 
\bibitem{Molina99}
Molina-Paris C., Visser M. {\sl Phys. Lett}~{\bf B 455} 90 1999
 
\bibitem{Tippett04}
Tippett B.K., Lake K. gr-qc/0409088
 
\bibitem{Szydlowski05}
Szydlowski M., Godlowski W., Krawiec A., Golbiak J. {\sl Phys Rev}~{\bf D72}  063504, 2005
 
\bibitem{Bojowald01}
Boyowald M. {\sl Phys. Rev Lett.}~{\bf 86}, 5227, 2001
 
\bibitem{Bojowald02}
Boyowald M. {\sl Phys. Rev Lett.}~{\bf 89}, 261301, 2002
 
\bibitem{Nojiri04a}
Nojiri S., Odintsov S.D. {\sl Phys. Lett.}~{\bf B595}, 1, 2004
 
\bibitem{Nojiri04b}
Elizalde E., Nojiri S., Odintsov S.D. {\sl Phys. Rev.}~{\bf D70}, 043539, 2004
 
\bibitem{Nojiri04c}
Nojiri S., Odintsov S.D. {\sl Phys. Rev.}~{\bf D70}, 103522,  2004
 
\bibitem{Tolman:1934}
Tolman R.C, {\sl Relativity Thermodynamics and Cosmology}
Oxford University Press, Oxford, 1934
 
\bibitem{Robertson:1933}
Robertson H.P. {\sl Rev. Mod. Phys.}{\bf 5}, 62 1933
 
\bibitem{Einstein:1931}
Einstein A. {\sl Berl. Ber.}  235 1931
 
\bibitem{Tolman:1931}
Tolman  R.C. {\sl Phys. Rev.}~{\bf 38} 1758 1931
 
\bibitem{Steinhardt:2002kw}
Steinhard P.J., Turok  N. {\sl Nucl. Phys. Proc. Suppl.}~{\bf 124} 38 2003
 
\bibitem{Shtanov:2002ek}
Shtanov Y, Sahni V. {\sl Class. Quantum Grav.}~{\bf 19}, L101 2002
 
\bibitem{Coule:2004qf}
Coule D.H., {\sl Class. Quantum Grav.} {\bf 22}, R125 2005
 
\bibitem{Salim:2005}
Salim J.~M., Perez~Bergliaffa S.~E., Souza .,
{\sl Class. Quantum Grav.}~{\bf 22} 975 2005
 
\bibitem{PintoNeto:2004wf}
Pinto~Neto ., {\sl Int. J. Mod. Phys.} {\bf D13}, 1419 2004
 
\bibitem{Barrow:2004ad}
 Barrow J.~D., Kimberly D., Magueijo J.,
{\sl Class. Quantum Grav.}~{\bf 21} 4289 2004
 
\bibitem{Bennett03}
Bennett C.~L. et~al. {\sl Astrophys. J. Suppl.}~{\bf 148} 1,  2003
 
\bibitem{Barris03}
Barris B.~J. et~al.{\sl Astrophys. J}~{\bf 602}, 571, 2004.
 
\bibitem{Chou04}
Choudhury T.R., Padmanabhan T. {\sl Astron. Astrophys.} {\bf 429}, 807, 2005
 
\bibitem{Knop03}
Knop R.~A. et~al. {\sl Astrophys.J}~{\bf 598}, 102, 2003
 
\bibitem{Riess01}
Riess A., Nugent P.E., Gilliland R.L., et al., {\sl Astrophys. J.}~{\bf 560},
49, 2001
 
\bibitem{Riess04}
Riess A.~G. et~al. {\sl Astrophys. J.}~{\bf 607}, 665, 2004
 
\bibitem{Tonry03}
Tonry J. L., et al., {\sl Astrophys. J.}~{\bf 594}, 1, 2003
 
\bibitem{Williams03}
Williams B.~F. et~al., 2003 astro-ph/0310432
 
\bibitem{Freese02}
Freese K.,  Lewis, M., {\sl Phys. Lett. B} {\bf 540}, 1, 2002
 
\bibitem{Zhu03}
Zhu Z.-H., Fujimoto M.-K., {\sl Astrophys. J.}~{\bf 585}, 52, 2003
 
\bibitem{Sen03}
Sen S., Sen A. A. {\sl Astrophys. J.}~{\bf 588}, 1, 2003
 
\bibitem{Liddle04}
Liddle A.R., astro-ph/0401198 2004
 
\bibitem{Akaike74}
Akaike H., {\sl IEEE Trans. Auto Control} 19, 716, 1974
 
\bibitem{Schwarz:1978}
Schwarz G., Annals of Statistics 5  461, 1978
 
\bibitem{Parkinson:2005}
Parkinson D., Tsujikawa S., Basset B., Amendola L. {\sl Phys Rev}~{\bf D71} 063524, 2005
 
\bibitem{Jeffreys:1961}
Jeffreys H., Theory of Probability, 3rd Edition, Oxford University Press,
 Oxford, 1961
 
\bibitem{Mukherjee:1998wp}
Mukherjee S., Feigelson E.~D., Babu G.~J., Murtagh F., Fraley C., Raftery A.,
{\sl Astrophys. J.}~{\bf 508}  314, 1998
 
\bibitem{Stromgren37}
Str{\"o}mgren B. {\sl Exapt. Ery. Naturw.}~{\bf 16} 465, 1937
 
\bibitem{Szydlowski04}
Szydlowski M., Maciejewski A. {\sl J. Phys.}~{bf A37}, 1, 2004
 
\end{thebibliography}
\end{document}